\documentclass[12pt,epsf,epsfig,psfig]{article}
\usepackage{graphicx}
\usepackage{epsfig}
\def\J{$J/\psi$}

\def\X{$\chi_c$}

\def\P{$\psi'$}

\def\C{c{\bar c}}

\def\e{\epsilon}

\def\be{\begin{equation}}
\def\ee{\end{equation}}

\def\lsim{\raise0.3ex\hbox{$<$\kern-0.75em\raise-1.1ex\hbox{$\sim$}}}
\def\gsim{\raise0.3ex\hbox{$>$\kern-0.75em\raise-1.1ex\hbox{$\sim$}}}

\def\NP{{ Nucl.\ Phys.\ }}
\def\PL{{ Phys.\ Lett.\ }}
\def\PR{{ Phys.\ Rev.\ }}

\def\PRL{{ Phys.\ Rev.\ Lett.\ }}

\def\ZP{{ Z.\ Phys.\ }}
\def\EP{{Eur.\ Phys.\ J. C}}

\begin{document}

\vspace*{-1cm}

17.\ 5.\ 2004 \hfill BI-TP 2004/11 

\bigskip

\centerline{\Large \bf The SPS Heavy Ion Programme}

\vskip 0.6cm

\centerline{\large \bf Helmut Satz}

\vskip 0.3cm

\medskip

%\large

\centerline{Fakult\"at f\"ur Physik, Universit\"at Bielefeld}

\centerline{Postfach 100 131, D-33501 Bielefeld, Germany}
 
\centerline{and}

\centerline{Centro de F{\'i}sica das Interac{\c c}{\~o}es 
Fundamentais (CFIF)}

\centerline{Instituto Superior T\'ecnico, 
P-1049-001 Lisbon, Portugal}

\normalsize

\vskip 1cm

%\newpage

\noindent
{\bf \large States of Matter in QCD}

\bigskip

During the past fifty years, our concept of an elementary particle
has undergone a fundamental change. Today we understand hadrons as
bound states of quarks, and thus as composite. In strong interaction
physics, quarks have become the smallest building blocks of nature.
But the binding force between quarks confines them to their hadron,
which cannot be split into its quark constituents. In terms 
of individual existence, hadrons remain elementary. 

\medskip
 
This modification of our hadron picture has led to remarkable consequences
in strong interaction thermodynamics: at high temperature or density, 
hadronic matter must become a plasma of deconfined quarks and gluons. In 
return, strong interaction thermodynamics points out the limits of quark 
confinement: in a sufficiently hot or dense medium, quarks can become free. 

\medskip

Such high densities prevailed in the very early universe, until some
10$^{-5}$ seconds after the big bang; only then were quarks confined
to form hadrons. To create and study this primordial plasma in the
laboratory is one of the great challenges for current experimental 
physics. Various estimates indicate that the collision of two heavy 
nuclei at very high energy might indeed produce short-lived bubbles 
of deconfined matter. CERN has played a vital role in initiating the 
experimental use of heavy ion collisions to search for the quark-gluon plasma, 
and it has in the past two decades provided fundamental contributions 
which today form the basis of our present understanding of the field.  

\medskip

\begin{figure}[h]
\hskip0.7cm
\begin{minipage}[t]{6cm}
\hskip-0.3cm \epsfig{file=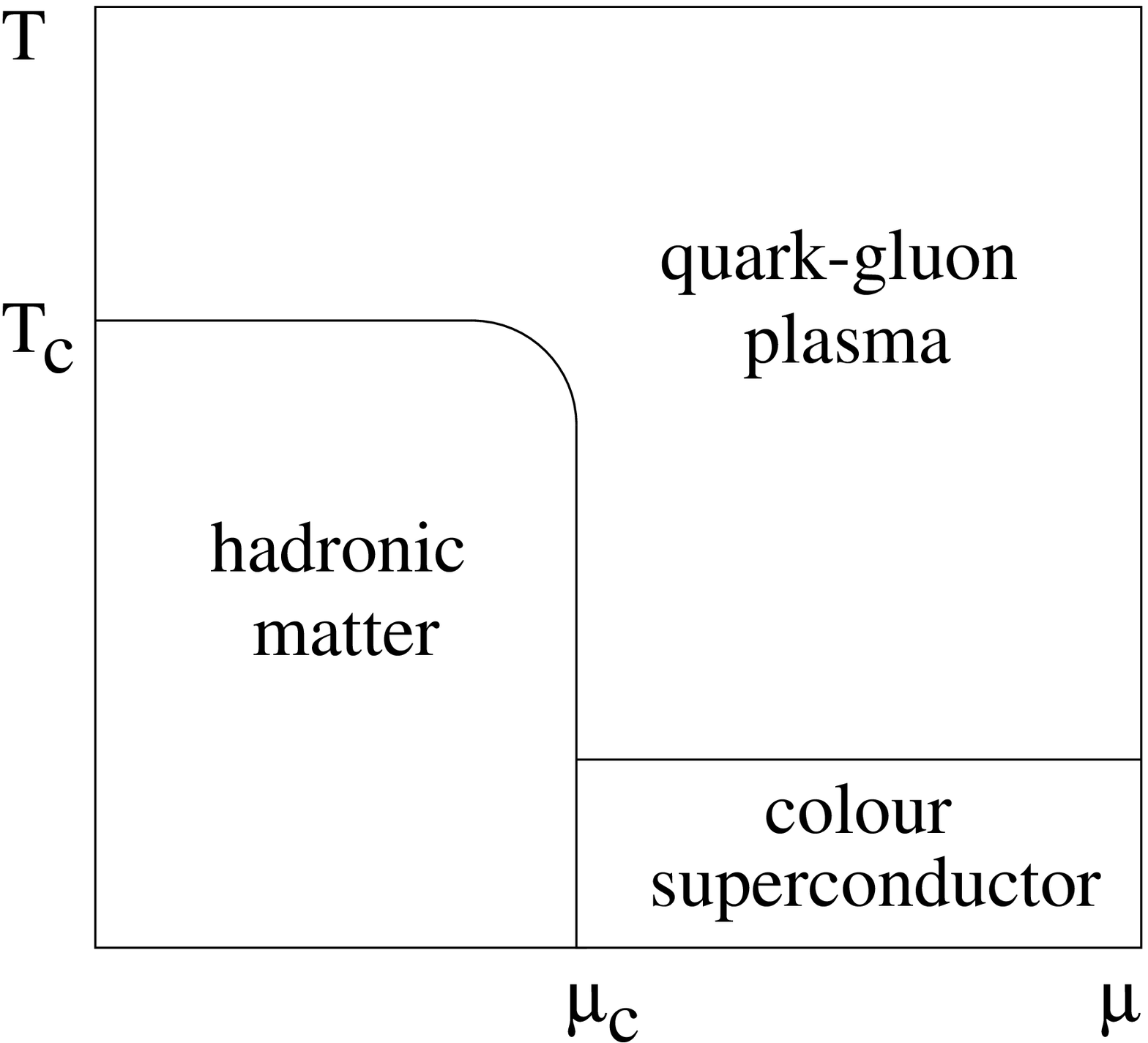,width=6.1cm,height=4.7cm}
\caption{Phase diagram of QCD}
\label{phase}
\end{minipage}
\hspace{2.2cm}
\begin{minipage}[t]{6cm}
\hskip-0.8cm \epsfig{file=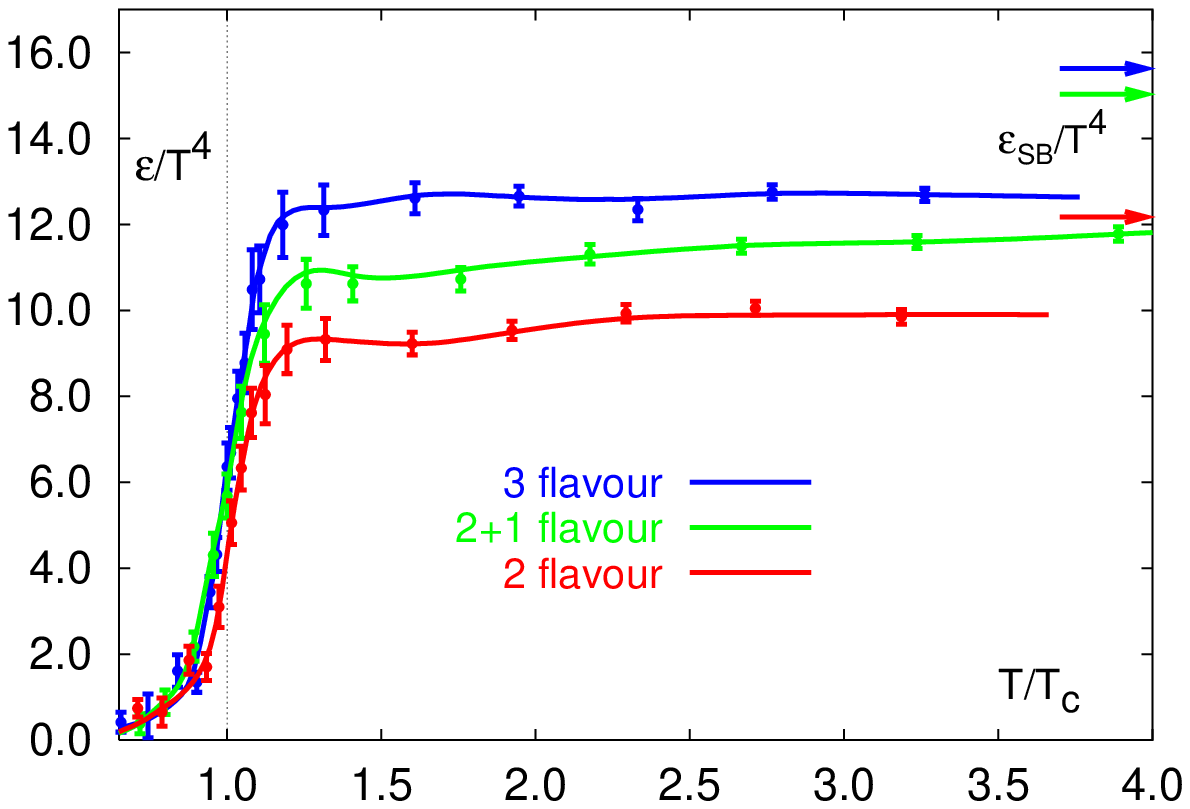,width=7.1cm}
\caption{Energy density in QCD} 
\label{ledens}
\end{minipage}
\end{figure}

Conceptually, the thermodynamics of strongly interacting matter 
leads to three forms of critical behaviour. 
\begin{itemize}
\vspace*{-0.2cm}
\item{In QCD, hadrons are dimensionful colour-neutral bound states of 
pointlike coloured quarks and gluons. Hadronic matter, consisting 
of colourless constituents of hadronic dimension, can therefore turn 
into a quark-gluon plasma of pointlike coloured quarks and gluons.
This deconfinement transition is the QCD counterpart of the 
insulator-conductor transition in atomic matter.}
\vspace*{-0.3cm}
\item{In vacuum, quarks dress themselves with gluons to form the constituent 
quarks that make up hadrons. As a result, the bare quark mass $m_q \sim 0$
is replaced by a constituent quark mass $M_q \sim 300$ MeV. In a
hot medium, this dressing melts and $M_q \to 0$. Since the QCD
Lagrangian for $m_q=0$ is chirally symmetric, $M_q \not= 0$ implies
spontaneous chiral symmetry breaking. The quark mass shift $M_q \to 0$ 
thus corresponds to chiral symmetry restoration.} 
\vspace*{-0.3cm}
\item{A third type of transition would set in if the attractive
interaction between quarks in the deconfined phase produces 
coloured bosonic diquarks, the Cooper pairs of QCD.
These diquarks can then condense at low temperature to form a colour
superconductor. Heating will dissociate the diquark pairs and turn
the colour superconductor into a normal colour conductor, the quark-gluon
plasma.}
\end{itemize}
\vspace*{-0.2cm}
\noindent
With the baryochemical potential $\mu$ as a measure for the
baryon density of the system, we thus expect the phase diagram 
of QCD to have the schematic form shown in Fig.\ \ref{phase}.

\medskip

Based on the QCD Lagrangian as dynamics input, the thermodynamics of 
strongly interacting matter is in principle fully specified, and at 
least for vanishing overall baryon density, finite temperature 
lattice QCD provides today quite detailed predictions \cite{karsch}. 
We here briefly summarize the most important features.

\medskip

The energy density of an ideal gas of massless pions is
\be
\e_h = 3 {\pi^2 \over 30} T^4 \simeq T^4,
\ee
while an ideal gas of massless quarks (for $N_f=2$) and gluons
gives
\be
\e_q = 37 {\pi^2 \over 30} T^4 \simeq 12 ~T^4.
\ee
Deconfinement thus produces a sudden increase in energy density, 
corresponding to the latent heat of deconfinement \cite{Celik}. 
This behaviour is in fact found in lattice QCD \cite{Biele1},
as shown in Fig.\ \ref{ledens}. For two light quark species, as well 
as for two light and one heavy species, the transition temperature becomes 
$T_c \simeq 175 \pm 15$ MeV, and the resulting energy density at 
deconfinement becomes $\e(T_c) \simeq 0.3 - 1.3$ GeV/fm$^3$. The 
abrupt change of behaviour of the energy density can be related 
directly to deconfinement and chiral symmetry restoration.

\medskip

Deconfinement is specified by the Polyakov loop expectation 
value \cite{kuti}
\be
L \sim \exp\{-F_{Q \bar Q}/T\}
\ee
where $F_{Q \bar Q}$ denotes the free energy of a $Q \bar Q$ pair in the
limit of infinite separation. In the confinement regime, $F_{Q \bar Q}$
diverges and hence $L=0$; in a deconfined medium, colour screening makes 
the free energy finite and hence $L \not= 0$. Thus the change of behaviour 
of $L$ defines the deconfinement temperature $T_L$.

\medskip

Chiral symmetry restoration is determined by
the chiral condensate $\langle \bar \psi \psi \rangle \sim
M_q$, which measures the dynamically generated constituent quark mass
$M_q$. When $\langle \bar \psi \psi \rangle \not= 0$, the chiral symmetry 
of the Lagrangian is spontaneously broken, and when 
$\langle \bar \psi \psi \rangle = 0$, it is restored. Hence
the change of behaviour of $\langle \bar \psi \psi \rangle$
defines the chiral symmetry restoration point $T_{\chi}$.

\medskip

In detailed lattice studies \cite{Biele2} it is shown that the 
two transitions clearly occur at the same temperature: at $\mu=0$, 
chiral symmetry restoration and deconfinement coincide. We can thus 
conclude that QCD predicts for $\mu=0$ one thermal transition
from hadronic matter to a quark-gluon plasma. For $N_f=2$ or 2+1, it
occurs at $T_c\simeq 175$ MeV; at this temperature, chiral symmetry is
restored, deconfinement sets in, and the energy density increases quite
suddenly by the ``latent heat" of deconfinement.

\medskip

\begin{figure}[h]
%\hskip0.7cm
\begin{minipage}[t]{7cm}
\hskip0.3cm 
\epsfig{file=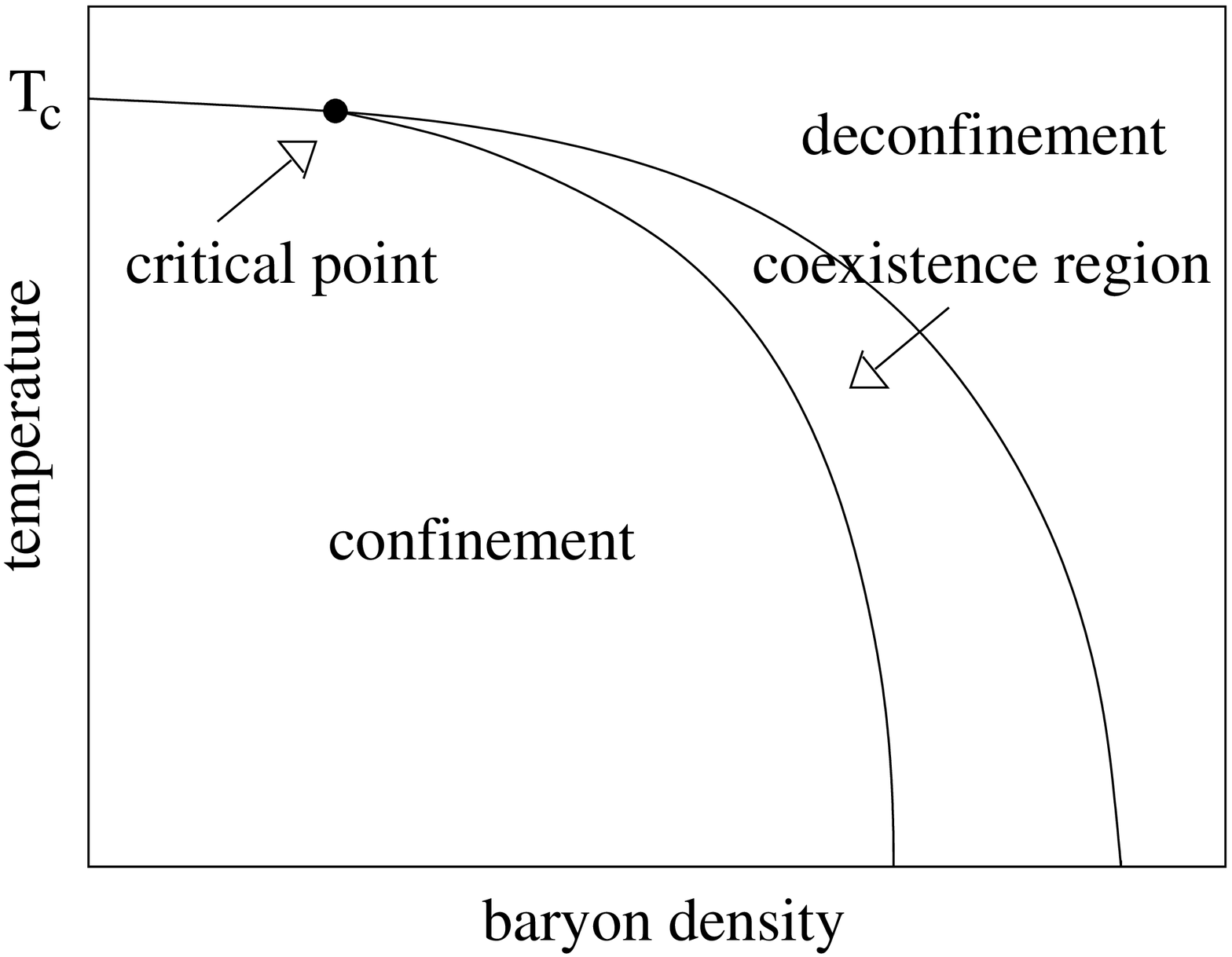,width=6.5cm,height=4.7cm}
\caption{Phase diagram as function of baryon density}
\label{phase-d}
\end{minipage}
\hspace{1.8cm}
\begin{minipage}[t]{6.5cm}
\vspace{-4.8cm}
\epsfig{file=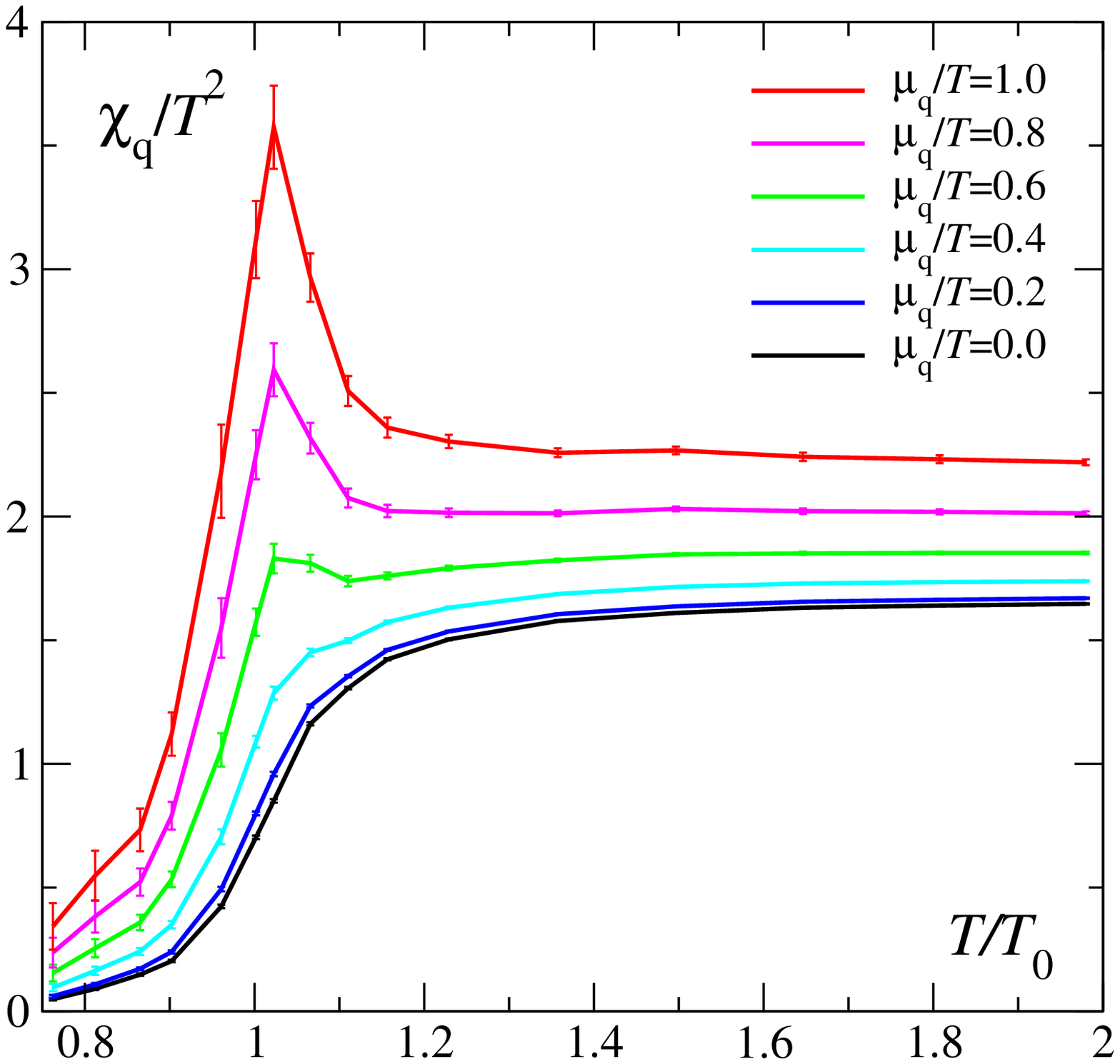,
width=6.5cm,height=4.6cm}
\vskip0.2cm
\caption{Baryon susceptibility $\chi_q$ vs.\ temperature} 
\label{fluc}
\end{minipage}
\end{figure}

The nature of the transition has been a subject of much attention by
theorists, but so far, it is not fully clarified, since it depends
quite sensitively on the baryon density as well as on $N_f$ and $m_q$. 
For a theory with one heavier and two light quarks, one expects \cite{Fodor}
non-singular behaviour (rapid cross-over, perhaps percolation) 
in a region between $0 \leq \mu < \mu_t$, a critical point at $\mu_t$, 
and beyond this a first order transition (see Fig.\ \ref{phase-d}). 
Recent lattice calculations in a theory with two quarks of finite mass
provide some support for such behaviour; as shown in Fig.\ \ref{fluc}, 
the baryon density fluctuations develop a pronounced peak with increasing 
baryochemical potential, which might indicate the approach to a nearby 
critical point \cite{bar-fluc}.

\medskip

\medskip

\noindent
{\bf \large High Energy Nuclear Collisions}

\bigskip

\begin{figure}[htb]
\centerline{\psfig{file=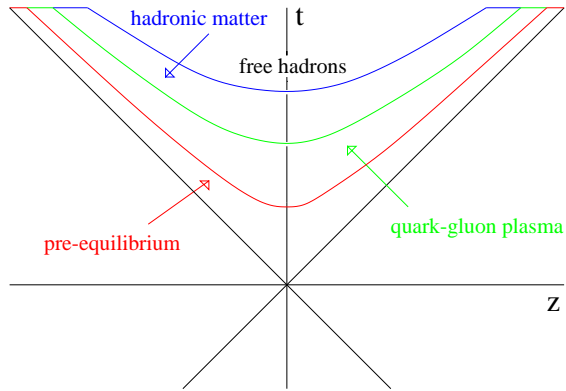,width=7.5cm}}
%\vskip0.5cm
\caption{Expected evolution of a nuclear collision}
\label{evo}
\end{figure}

The heavy ion programme at the CERN-SPS was created with the aim to 
produce the quark-gluon plasma in the laboratory and to study both the 
deconfinement transition and the new 
deconfined state of matter. Starting from the non-equilibrium configuration 
of the two colliding nuclei, the evolution of the collision was assumed to 
have the form illustrated in Fig.\ \ref{evo}. After the collision, there 
is a short pre-equilibrium stage, in which the primary partons of the 
colliding nuclei interact, multiply and then thermalize to form a 
quark-gluon plasma. This then expands, cools and hadronizes. 

\medskip

In recent years, the effect of pre-equilibrium conditions on deconfinement 
have been studied in more detail; in particular, it now appears conceivable
that nuclear collisions lead to a specific form of deconfinement without
ever producing a thermalized plasma of quarks and gluons. We shall return 
to these aspects later and here address first the probes proposed to study 
the different stages and properties of a thermal medium.

\medskip

The initial energy density of the produced medium at the time of 
thermalization was estimated by~\cite{Bjorken83}
\be
\e = \left({dN_h \over dy}\right)_{y=0} {w_h \over \pi R_A^2 \tau_0},
\label{2.1}
\ee
where $(dN_h/dy)_{y=0}$ specifies the number of hadrons
emitted per unit rapidity at mid-rapidity and $w_h$ their average
energy. The initial volume is determined in the transverse
nuclear size (radius $R_A$) and the formation time $\tau_0 \simeq
1$ fm of the thermal medium.

\medskip

The determination of the nature of the hot initial phase required 
deconfinement signatures. It was argued that in a hot quark-gluon plasma,
the J/$\psi$ would melt through colour screening~\cite{MS}, so that QGP 
formation should lead to a suppression of J/$\psi$ production in nuclear 
collisions. Similarly, the QGP was expected to result in a higher 
energy loss for a fast passing colour charge than a hadronic medium, so 
that increased jet quenching~\cite{jets} should also signal deconfinement.

\medskip

The temperature of the produced medium, in the confined as well as in
the deconfined phase, was assumed to be observable
through the mass spectrum of thermal dileptons and the
momentum spectrum of thermal photons~\cite{Shuryak,Keijo}. The
observation of thermal dilepton/photon spectra would also indicate
that the medium was indeed in thermal equilibrium. 

\begin{figure}[htb]
\vskip-2.5cm
\centerline{\psfig{file=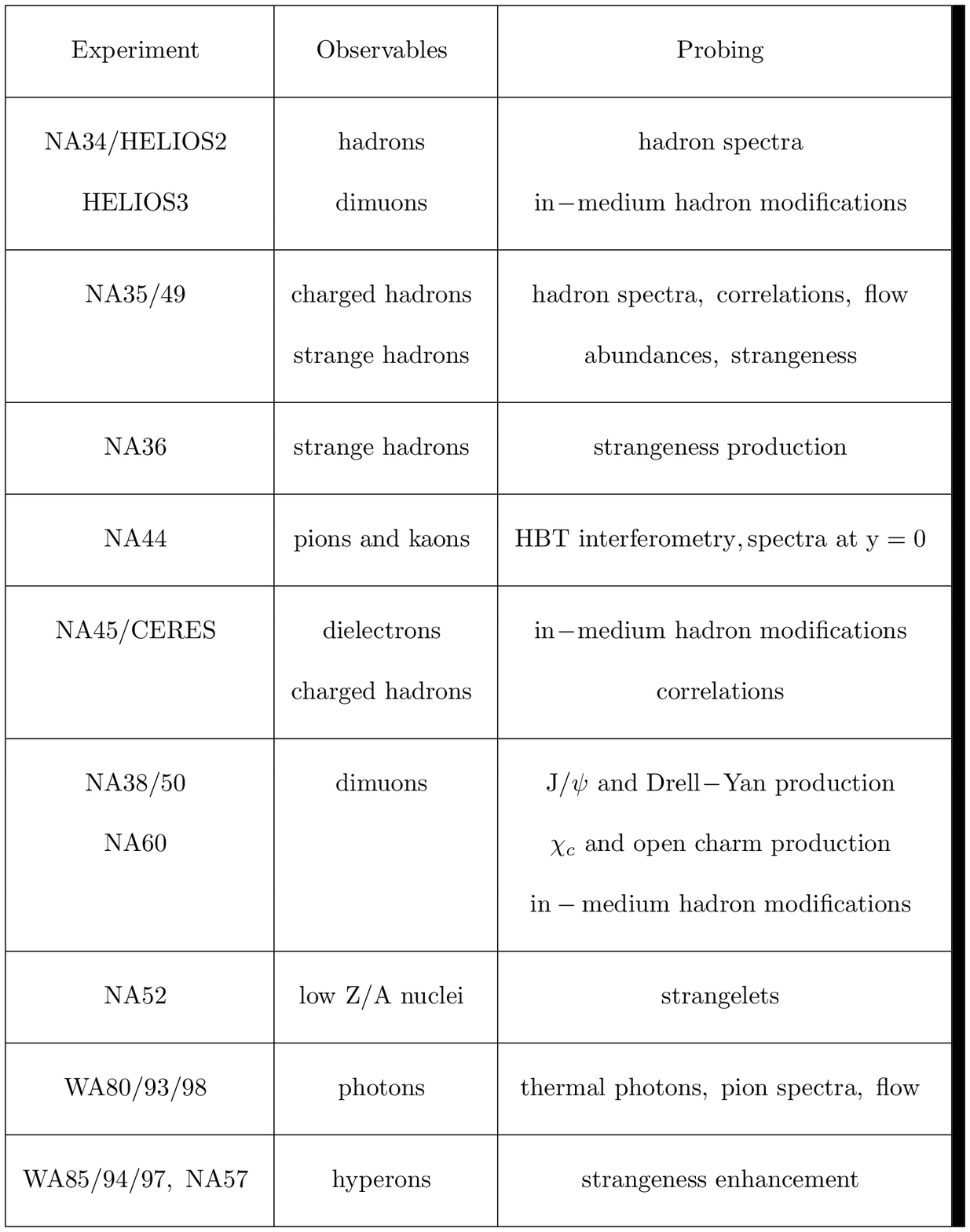,width=15cm}}
\vskip-3.5cm
\centerline{Table 1: Heavy ion experiments at the CERN-SPS}
\end{figure}

\medskip

The behaviour of sufficiently short-lived resonances, in particular the
dilepton decay of the $\rho$, was considered as a viable tool to study
the hadronic medium in its interacting stage and thus provide information 
on the approach to chiral symmetry restoration~\cite{rho-chiral}.

\medskip

The expansion of the hot medium was thought to be measurable through
broadening and azimuthal anisotropies of hadronic transverse momentum
spectra (flow)~\cite{flow}. The size and age of the source at freeze-out was
assumed to be obtainable through Hanbury-Brown--Twiss (HBT)
interferometry based on two-particle correlations~\cite{HBT}.  It was
expected that increasing the collision energy would increase the
density and hence the expansion of the produced medium, so that the
HBT radii should grow with increasing $\sqrt s$ \cite{Stock-A}. 

\medskip

The final interacting hadronic medium had been discussed in terms of an
ideal resonance gas, which at vanishing overall baryon density would 
provide the relative abundances of all hadron species in terms of just 
one parameter, the limiting temperature of hadronic matter \cite{Hagedorn}.
The species abundances in elementary hadronic interactions follow such 
a pattern \cite{Becattini}, but with an overall reduction of strangeness 
production. Nuclear collisions, if leading to the formation of a hot 
quark-gluon plasma with a thermal density of strange quarks and antiquarks, 
were expected to remove this reduction and thus result in enhanced strangenes 
production in comparison to $pp$ interactions~\cite{Rafelski}. -- The 
formation of strange baryonic matter (`strangelets') was also considered
\cite{strangelet}.

\medskip

In order to address the features just outlined, CERN started an
extensive experimental programme, as summarized in Table 1. Since
the ion injector available in 1986 was restricted to nuclei with
equal numbers of protons and neutrons, the initial programme used
$^{16}O$ and $^{32}S$ beams of $P_{Lab}/A \simeq 200$ GeV/c
on different heavy targets.  Subsequently an injector was constructed
to accommodate arbitrarily heavy nuclei; with its help, the use of 
$^{208}Pb$ beams with $P_{Lab}/A \simeq 160$ GeV/c
started in 1995. We now summarize the main results
obtained by the programme so far; one experiment (NA60) is continuing
with $^{115}In$ beams at least up to the year 2005.

\bigskip

\noindent
{\bf \large Experimental Results}

\bigskip

\begin{figure}[h]
\hskip0.5cm
\begin{minipage}[t]{6.5cm}
\hskip-0.3cm \epsfig{file=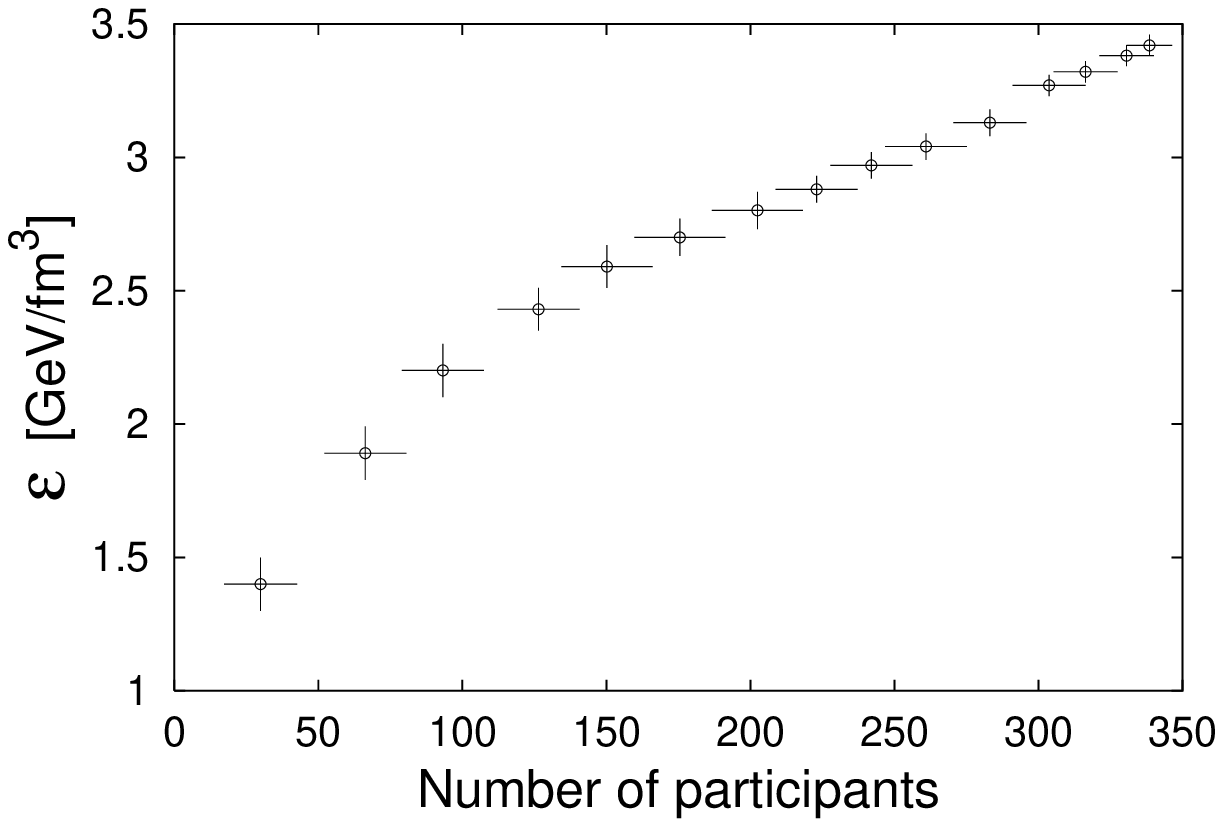,width=7cm, height=5.2cm}
\caption{Energy density in Pb-Pb collisions \cite{NA50}.}
\label{edens}
\end{minipage}
\hspace{1.5cm}
\begin{minipage}[t]{6.5cm}
\vspace{-5.4cm}
\hskip-0.5cm \epsfig{file=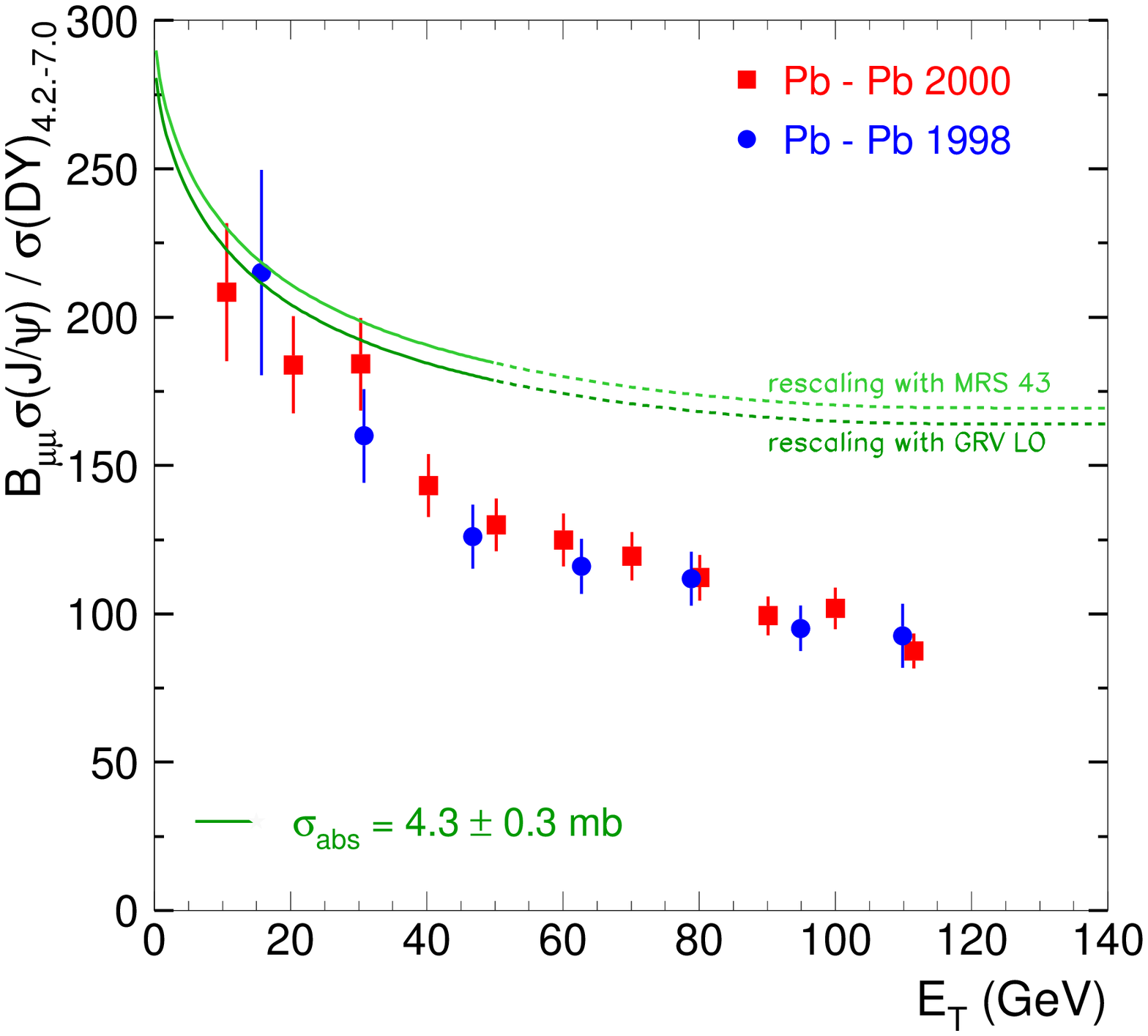,width=7.7cm, height=6.0cm}
\vskip-0.8cm
\caption{Ratio of \J~to Drell-Yan production in Pb-Pb 
collisions \cite{NA50}.}
\label{na50} 
\end{minipage}
\end{figure}

The initial energy density, as specified by the Bjorken estimate,
Eq.\ (\ref{2.1}), was measured in almost all SPS experiments. In
Fig.\ \ref{edens}, we show $\e$ as function of centrality, determined
by the number of participant nucleons~\cite{NA50}; it covers the
range from somewhat above 1 to almost 3.5~GeV/fm$^3$ and thus 
reaches well above the deconfinement value. 

\medskip

J/$\psi$ production was found to be suppressed in O-U, S-U and then
Pb-Pb collisions; the suppression always increases with
centrality~\cite{NA38/50}.  In p-A collisions, it was observed
that already normal nuclear matter leads to reduced charmonium
production. Extrapolating this `normal' suppression to nucleus-nucleus
interactions is enough to account for the observed yields up to
central S-U collisions. Peripheral Pb-Pb collisions also follow the 
normal pattern; then, with increasing centrality, there is a pronounced 
onset of a further `anomalous' suppression \cite{NA50}, as shown in 
Fig.~\ref{na50}. 

\medskip

\begin{figure}[h]
\hskip0.5cm
\begin{minipage}[t]{6.8cm}
\hskip-0.3cm \epsfig{file=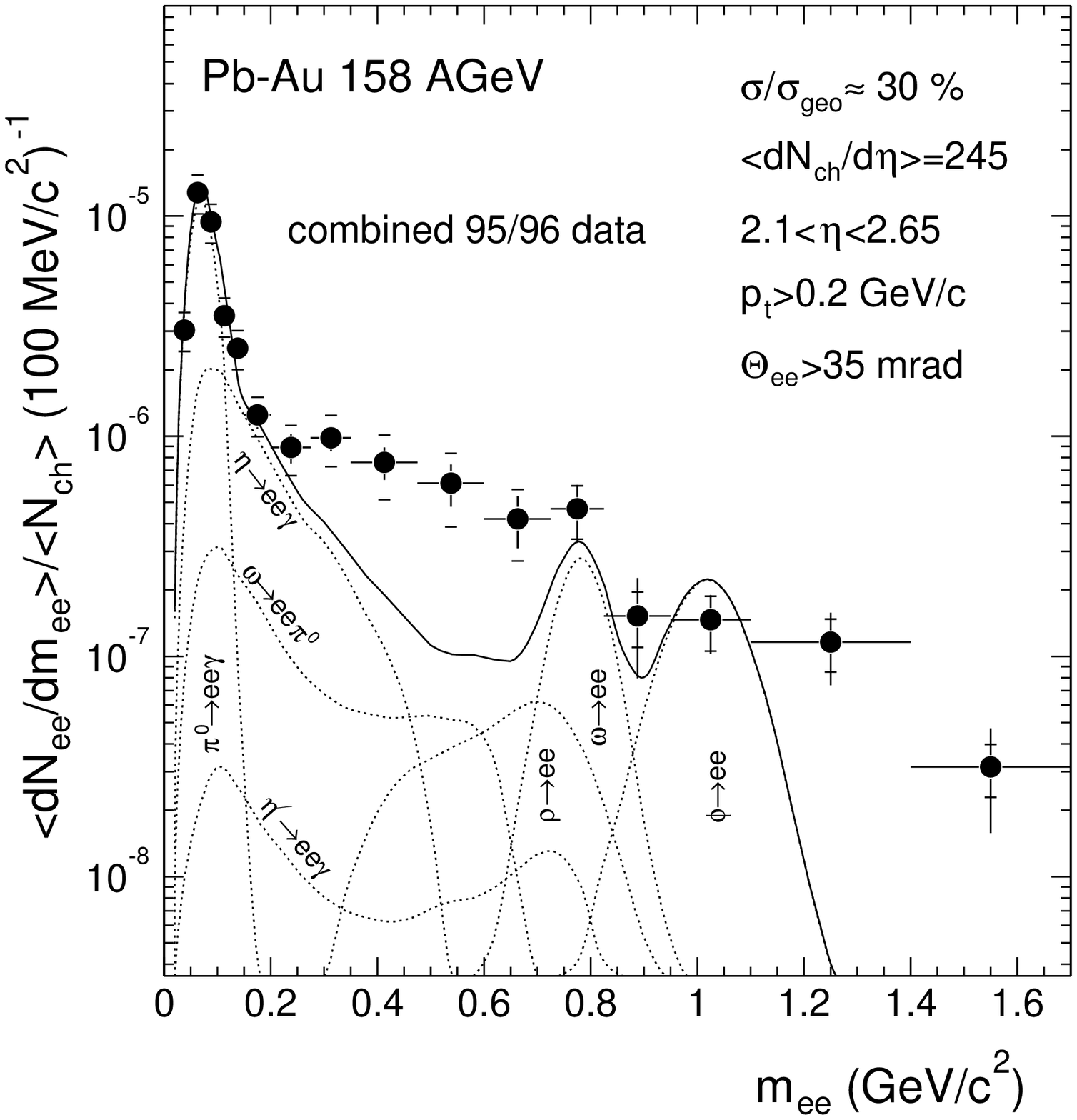,width=7cm, height=5cm}
\vskip-0.2cm
\caption{Dilepton production 
compared to the expected yield from known
hadronic sources \cite{Ceres}.}
\label{rho}
\end{minipage}
\hspace{1.8cm}
\begin{minipage}[t]{6.5cm}
\vspace{-5.3cm}
\hskip-0.5cm \epsfig{file=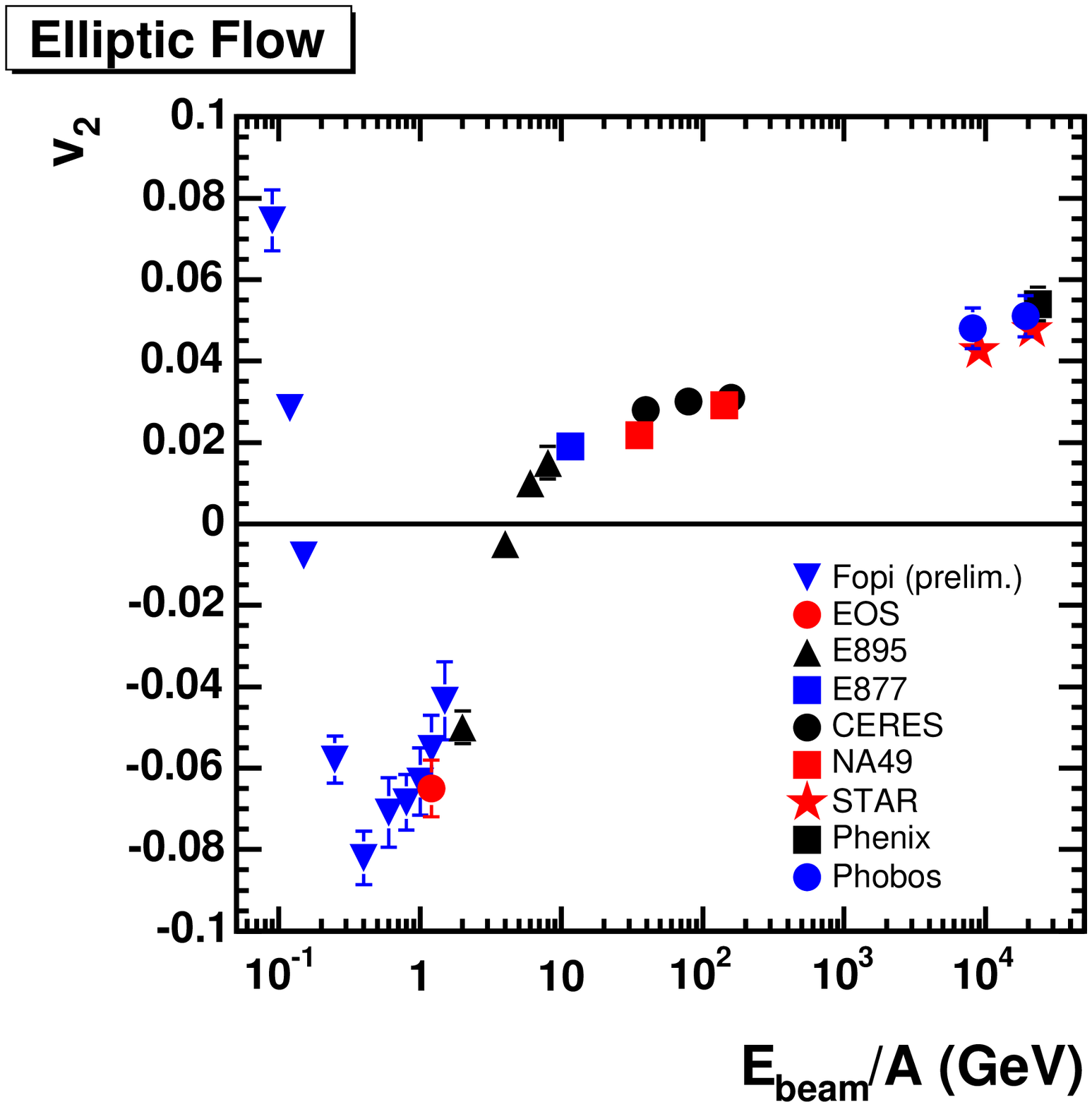,width=7.1cm, height=5.8cm}
\vskip-0.67cm
\caption{Elliptic flow at different beam energies \cite{stock}.}
\label{ellip} 
\end{minipage}
\end{figure}

In $pp$ and $pA$ collisions, the dilepton mass spectrum in the region 
around and below the $\rho$ peak is well reproduced by the yield from 
known hadronic sources. In $AA$ collisions, it was found to differ 
considerably from this expected yield \cite{Ceres,Masera}, indicating 
in-medium resonance modifications (Fig.~\ref{rho}). This `low mass dilepton 
enhancement' is observed in S-U and Pb-Pb collisions, and for the latter 
at beam energies of 40~GeV as well as of 160~GeV \cite{Specht}.

\medskip

Similarly, some photon excess over the expected normal hadronic decay 
yield has been reported~\cite{photondata}, as well as an excess of 
dileptons in the mass range between the $\phi$ and the 
J/$\psi$~\cite{hm-dileptons}. 

\medskip

The broadening of transverse momentum spectra, expected as consequence
of transverse flow, was observed in the predicted form, increasing
with increasing hadron mass. Moreover, transverse momentum spectra also 
showed the azimuthal anisotropy (elliptic flow) predicted for non-central 
collisions. The behaviour shown in Fig.~\ref{ellip} indicates that at 
low collision energy, production is reduced by the presence of spectator
nucleons; at high energy, there is enhanced production in the direction 
of the higher pressure gradient as determined by the anisotropic interaction 
volume~\cite{stock}.

\medskip

\begin{figure}[htb]
\begin{minipage}[t]{6cm}
\hskip0.5cm
\epsfig{file=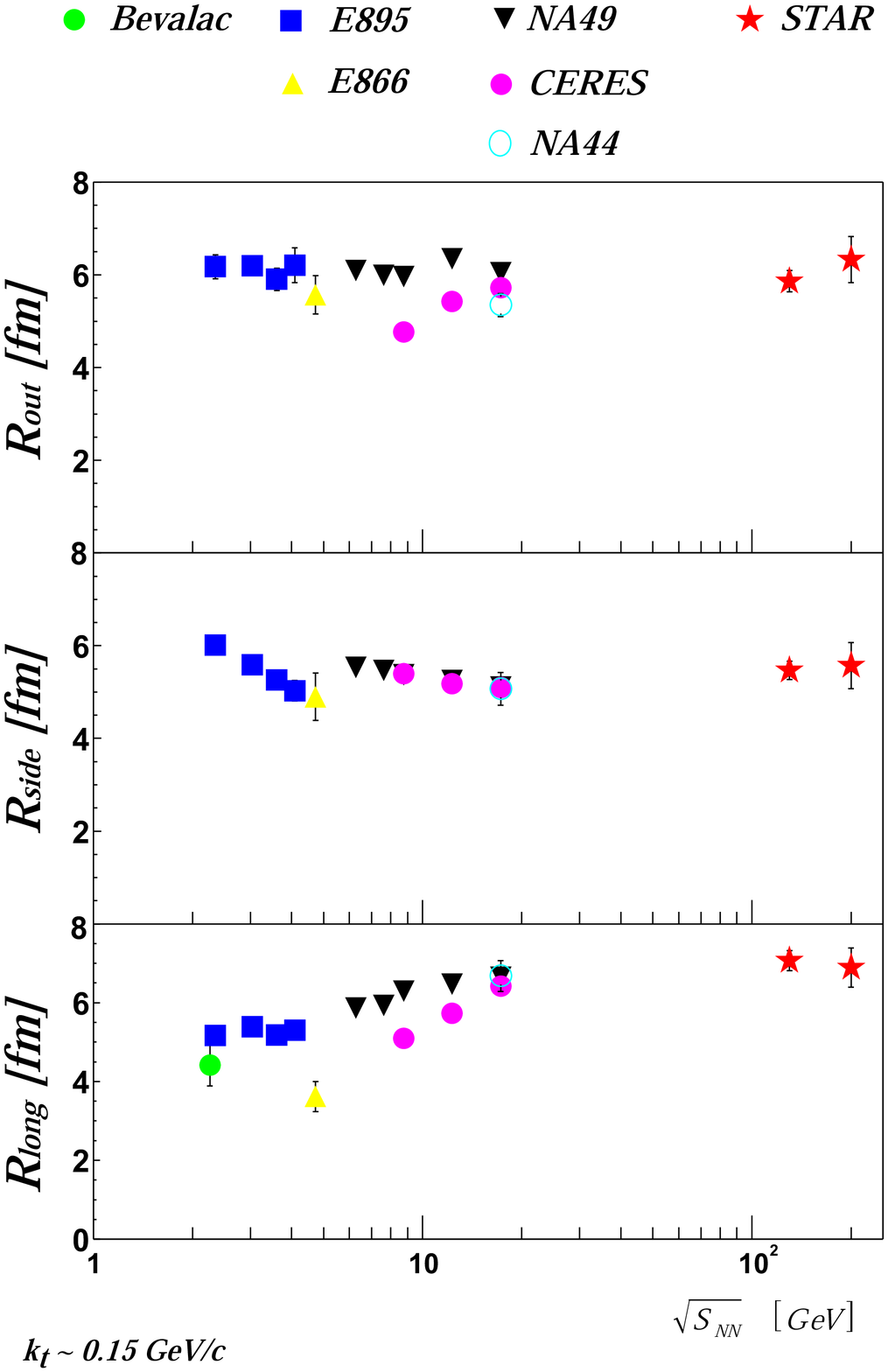,width=5cm,height=6.8cm}
\caption{HBT radii at different beam energies \cite{stock}.}
\label{HBT}
\end{minipage}
\hspace{0.7cm}
\begin{minipage}[t]{8.5cm}
\vspace*{-6.53cm}
\hskip-0.3cm
\epsfig{file=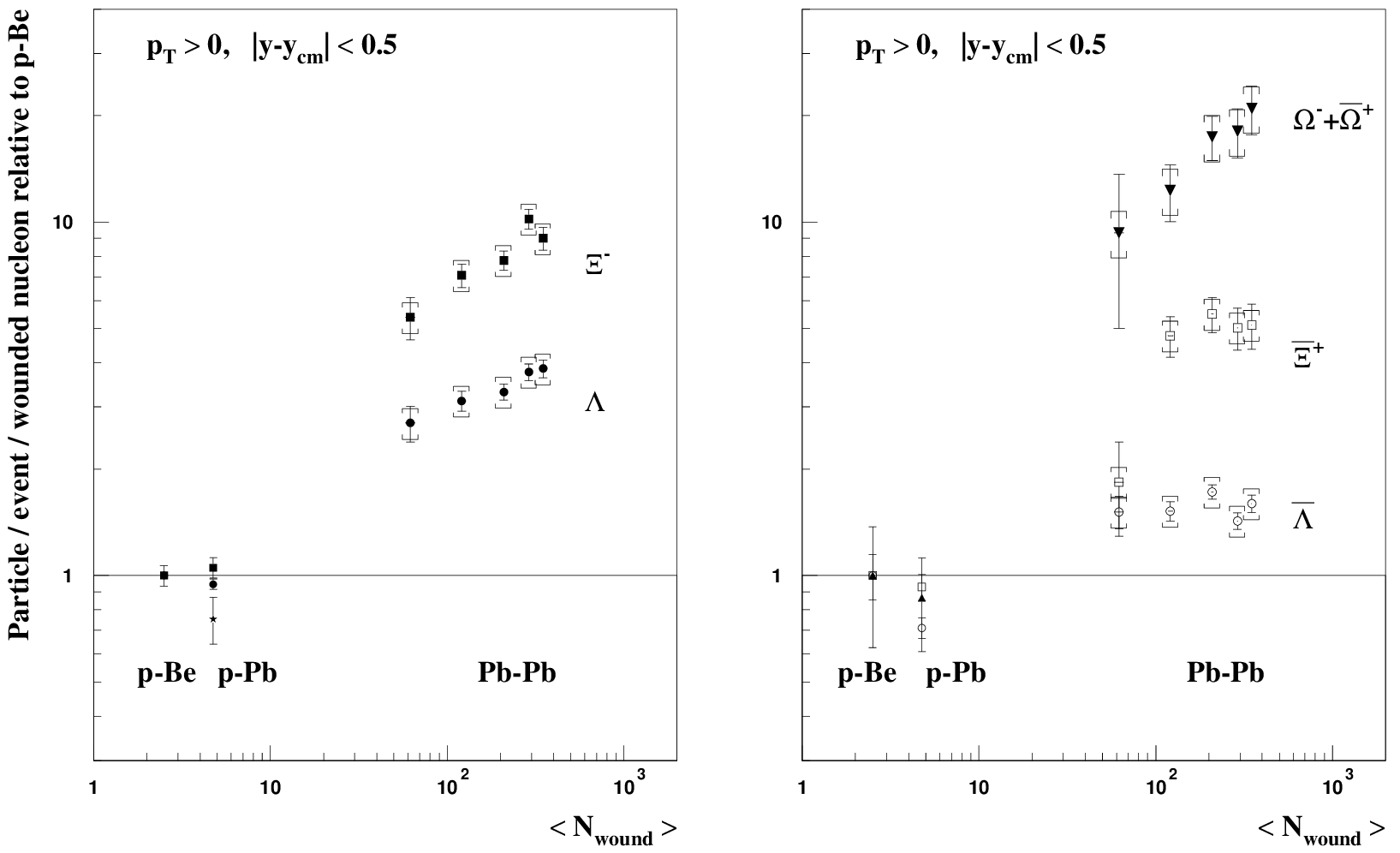,width=6.7cm,height=8.8cm,angle=-90}
\caption{Hyperon production as function of centrality, normalized
to p-Be results \cite{antinori}.}
\label{Antinori}
\end{minipage}
\end{figure}

In HBT correlation studies it is found that at all energies the source 
radii are essentially determined by those of the involved nuclei~\cite{stock};
the expected increasing source size was not observed. Thus one finds
for the transverse radii $R_{\rm side} \simeq R_{\rm out} \simeq 5-6$~fm 
for Au-Au/Pb-Pb collisions from AGS to SPS and on to RHIC, as seen in 
Fig.~\ref{HBT}. The approximate equality of $R_{\rm out}$ and $R_{\rm side}$ 
is another unexpected feature, since their difference should be a measure 
of the life-time of the emitting medium.

\medskip

The expected enhancement of strangeness production was indeed observed; 
the ratio of strange to non-strange hadrons in $AA$ collisions, compared
to the same ratio in $pp$ collisions, always increased significantly. In  
Fig.~\ref{Antinori} we show the most striking example, in which the production 
of hyperons is increased up to 10 times and more in comparison to $p-Be$ rates
~\cite{antinori}. -- No indication for strangelet production was 
found \cite{newmass}

\bigskip

\noindent
{\bf \large Conclusions, Questions, Outlook}

\bigskip

The wealth of experimental results obtained so far has clearly established 
that high energy nuclear interactions produce large-scale complex systems 
which show a number of specific collective effects and thus provide 
more than a superposition of independent nucleon-nucleon collisions.
Let us see what conclusions the individual observations lead to. 

\medskip

The essential issues for the CERN heavy ion programme were:
\begin{itemize}
\vspace*{-0.2cm}
\item{do the colliding nuclei produce a system of deconfined 
quarks and gluons,}
\vspace*{-0.2cm}
\item{can the produced system be described in thermal terms,}
\vspace*{-0.2cm}
\item{is there any rapid change of an observable, indicating critical
behaviour.}
\end{itemize}
\vspace*{-0.2cm}
The quark-gluon plasma of statistical QCD is a deconfined system in
thermal equilibrium. However, as we shall see shortly, pre-equilibrium 
studies suggest that in nuclear collisions, deconfinement and thermalization 
should be addressed as two distinct issues.

\medskip 

Hard probes meant to test the nature of the early medium can suffer 
nuclear effects at different points in the evolution of the collision. 
The presence of a nuclear target can modify the production of the probe, 
as seen in $pA$ collisions. Once formed, the probe can interact with 
the partons in the pre-equilibrium stage, and/or it can subsequently 
be effected by the QGP and the final hadronic medium. It is thus 
important to distinguish between initial nuclear effects from those 
due to different evolution stages of the medium formed by the collision. 

\medskip

\begin{figure}[h]
\centering
\begin{tabular}{cccccc}
\resizebox{0.2\textwidth}{!}{%
\includegraphics*{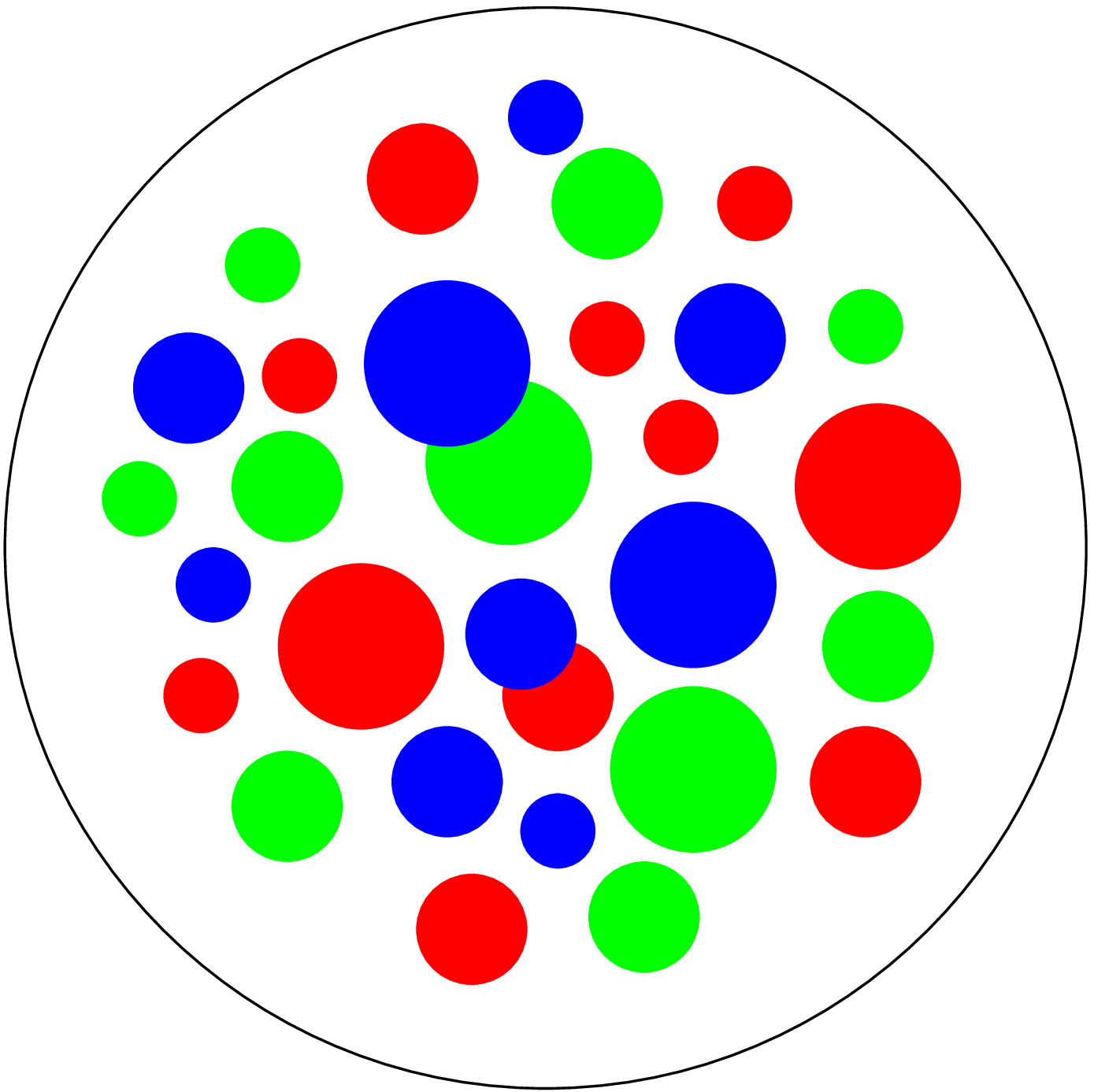}}
& & & & &
\resizebox{0.2\textwidth}{!}{%
\includegraphics*{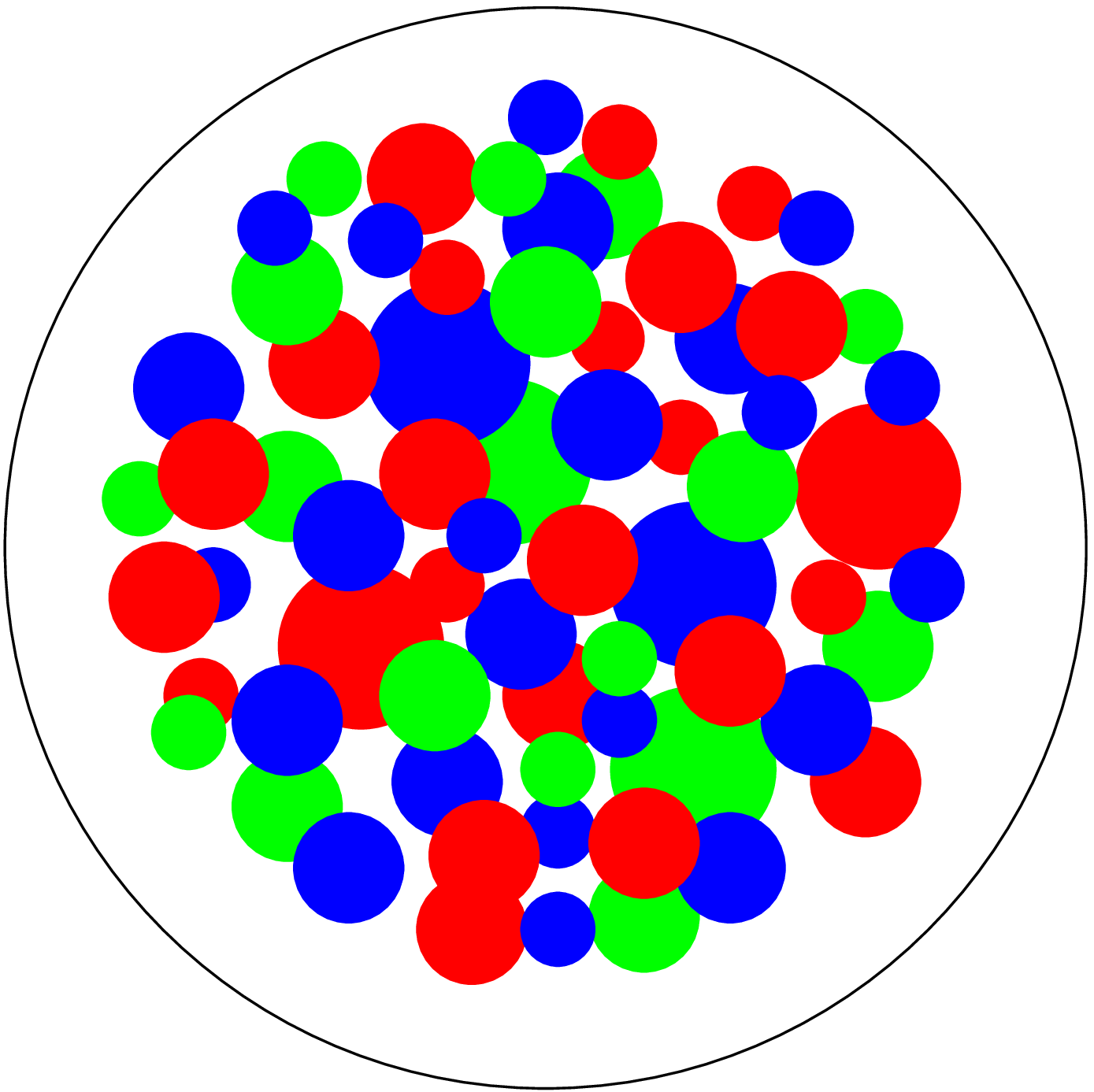}}
\end{tabular}
\caption{Parton distributions in the transverse plane of a
nucleus-nucleus collision}
\label{PP}
\vglue-4mm
\end{figure}

The partonic constituents in the initial state of a high energy nuclear 
collision are given by the parton distribution function of the
colliding nuclei. To produce a large-scale thermal system, partons from 
different nucleon-nucleon collisions have to undergo multiple interactions. 
In the center of mass initial state of a high energy collisions, the nuclei 
are strongly Lorentz contracted; the resulting parton distribution in the 
transverse collision plane is schematically illustrated in Fig. \ref{PP}. 
The transverse size of the partons is determined by their intrinsic 
transverse momentum, and the number of partons contained in a nucleon is 
known from deep inelastic scattering experiments. The density of partons 
increases with both $A$ and $\sqrt s$, and at some critical point, parton 
percolation occurs \cite{PP} and ``global'' colour connection sets in. In 
the resulting parton condensate, partons lose their independent existence 
and well-defined origin, so that this medium is deconfined, though not 
thermalized. In recent years, such partonic connectivity requirements 
(closely related to parton saturation) and the properties of a connected 
pre-thermal primary state (colour glass condensate) have attracted much 
attention \cite{PP,cgc}. 

\medskip

We thus have to determine if some specific observed behaviour is due 
to nuclear effects on the formation of the probe, to the parton condensate 
in the initial pre-equilibrium stage, or to the presence of a deconfined 
or confined thermal medium. The problem is particularly transparent in the 
case of \J~suppression.  

\medskip

As already mentioned, charmonium production in $pA$ collisions is reduced 
in comparison to that in $pp$ collisions, so that the presence of the 
nuclear target modifies the production of the probe. Once this `normal' 
suppression, shown by the solid line in Fig.\ \ref{na50}, is taken into 
account, any further `anomalous' suppression, as seen in central $Pb-Pb$ 
collisions, is then due to the presence of a produced medium. 

\medskip

\J~production in $pp$ collisions has shown that only about 60\% of the 
observed \J's are produced directly as $1S$ $\C$ states; the remainder 
comes from decay of the (larger) excited states \X~and \P. Now the 
effect of any medium on a charmonium state depends on the intrinsic scales 
of medium and probe. It is today known from finite temperature lattice 
QCD \cite{spectral} that in a QGP the higher excited states \X~and \P~are 
dissolved at approximately the deconfinement point, while the smaller 
ground state \J~survives up to significantly higher temperature. 
Present calculations give 1.5 - 2 $T_c$ for the dissociation point,
but do not yet allow calculations of the width of the state as function
of temperature. In a similar fashion, a pre-equilibrium parton condensate 
with a given resolution scale can dissociate only charmonia it can resolve. 
Here it is also found that the larger excited states are suppressed at 
the overall onset of parton percolation, while the ground state survives 
up to higher parton densities \cite{DFS}. In both cases we thus expect 
a two-step suppression pattern: first the \X~and \P~disappear, which 
suppresses the \J's from their decay, later the directly produced ground 
state is suppressed. In contrast, any suppression in a hadronic medium 
leads to a smooth variation without threshold \cite{Ramona}. 

\begin{figure}[h]
\begin{minipage}[t]{7.4cm}
\epsfig{file=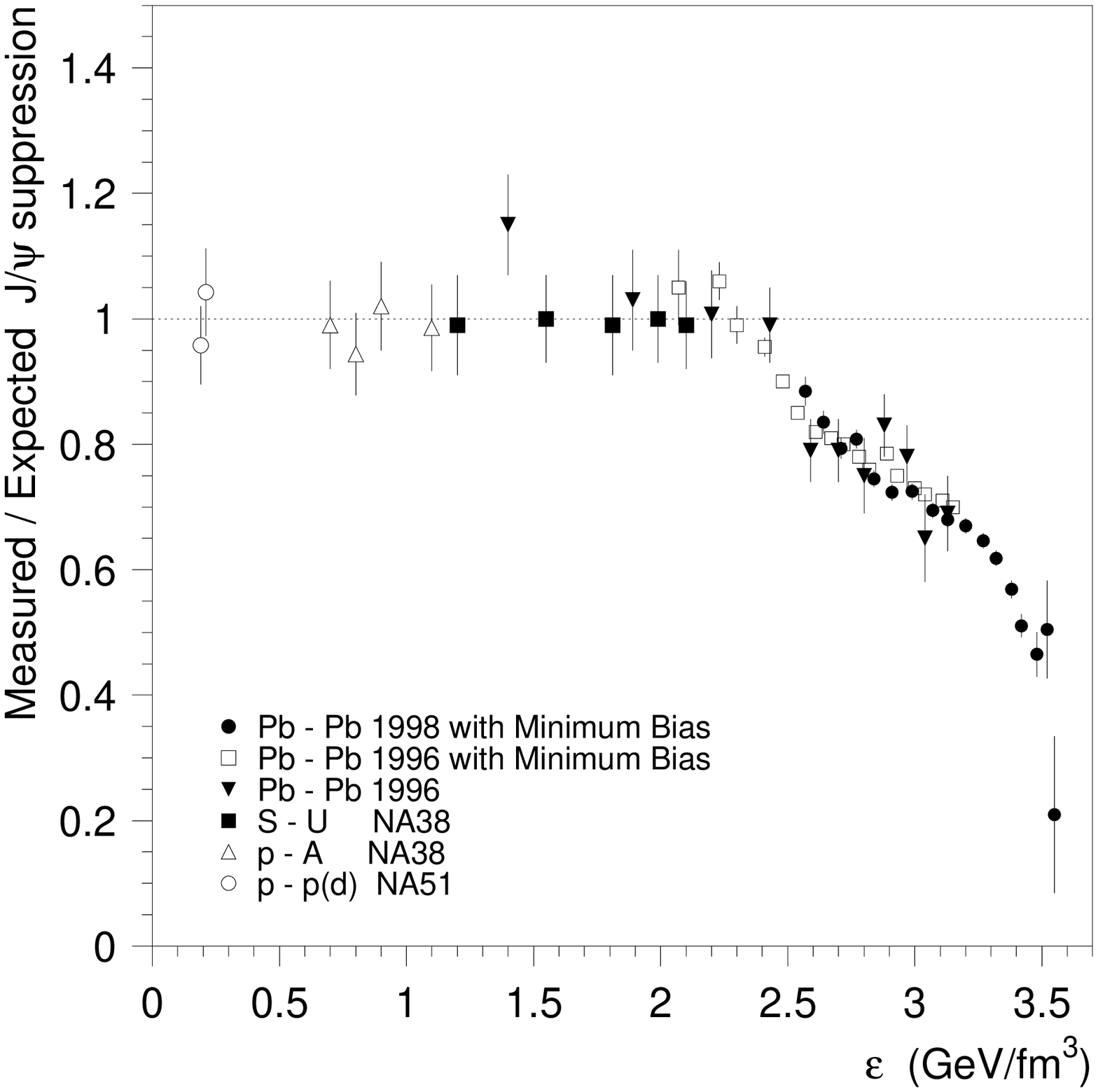,width=7cm, height=5.3cm}
\vskip0.1cm
\caption{\J~suppression vs.\ energy density \cite{NA50}.}
\label{e-psi}
\end{minipage}
\hspace{0.6cm}
\begin{minipage}[t]{7.4cm}
\vskip-5.5cm
%\hskip0.3cm
\epsfig{file=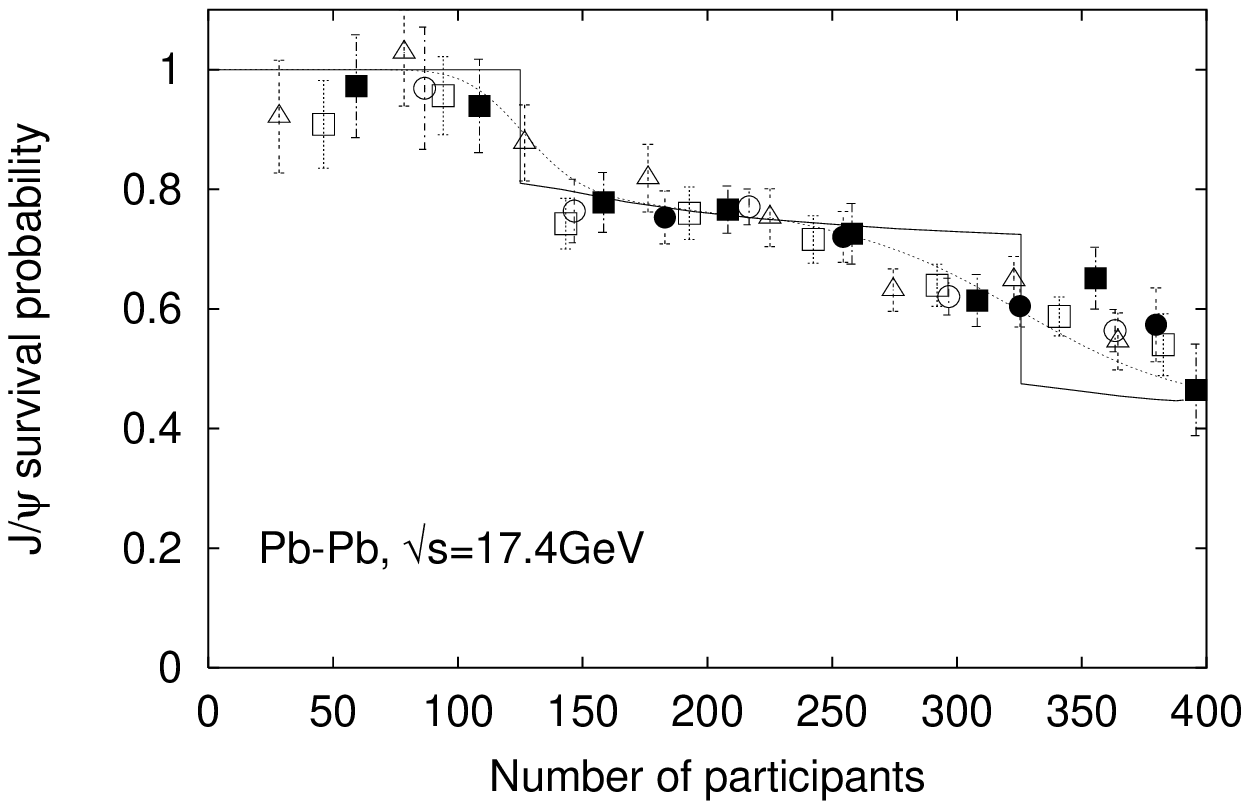,width=7.5cm, height=6cm}
\vskip-0.4cm
\caption{ \J~suppression and parton condensation \cite{DFS}.}
\label{supp} 
\end{minipage}
\end{figure}

The NA50 data \cite{NA50} indeed indicate a two-step pattern (see Fig.\ 
\ref{na50}), although the second step is less pronounced and has also been 
attributed to multiplicity fluctuations in central collisions \cite{b-o}.
Threshold energy density estimates, based on the Bjorken form (\ref{2.1})
with $\tau_0=1$ fm, give around 2 - 2.5  GeV/fm$^3$ for the initial onset 
of anomalous suppression; as seen in Fig.\ \ref{e-psi}, this does not agree 
with the lattice QCD result for deconfinement, 
$\e(T_c) \simeq 0.3 - 1.3 $ GeV/fm$^3$.
In contrast, the parton percolation threshold ($N_{\rm part} \simeq
125$) agrees well with the first step of the anomalous \J~suppression
(see Fig.\ \ref{supp}), removing \X~and \P~contributions; the second 
step is compatible with percolation of harder partons \cite{DFS}. 
Evidently the relevant variable for the threshold points is crucial, 
and this can be corroborated by experiments at different $A$ or $\sqrt s$,
since both enter in the parton density. 
Further confirmation and clarification of the threshold behaviour 
in \J~suppression is of decisive importance. It is the only 
clear onset of new behaviour, and hence the only indication for any form
of critical behavior seen in any heavy ion experiment, and it can only
be accounted for in terms of deconfinement. For this there
are two distinct possibilities: parton percolation in the pre-equilibrium 
stage, or colour screening in a thermalized stage. The final conclusion 
must necessarily come from experiment.

\medskip

All other phenomena studied at the SPS arise in the hadronic stage of 
the collision evolution and can thus be used to study the medium at
the hadronization transition and later on. Here the issue of thermal 
behaviour is of fundamental importance, since from our pre-hadronic
information it is so far not at all obvious that the collision of two heavy 
nuclei will produce systems which can be understood in terms of 
equilibrium thermodynamics. There are several observations which 
confirm that in the hadronic stage this is indeed the case.

\medskip

\begin{figure}[htb]
\begin{minipage}[t]{7cm}
\epsfig{file=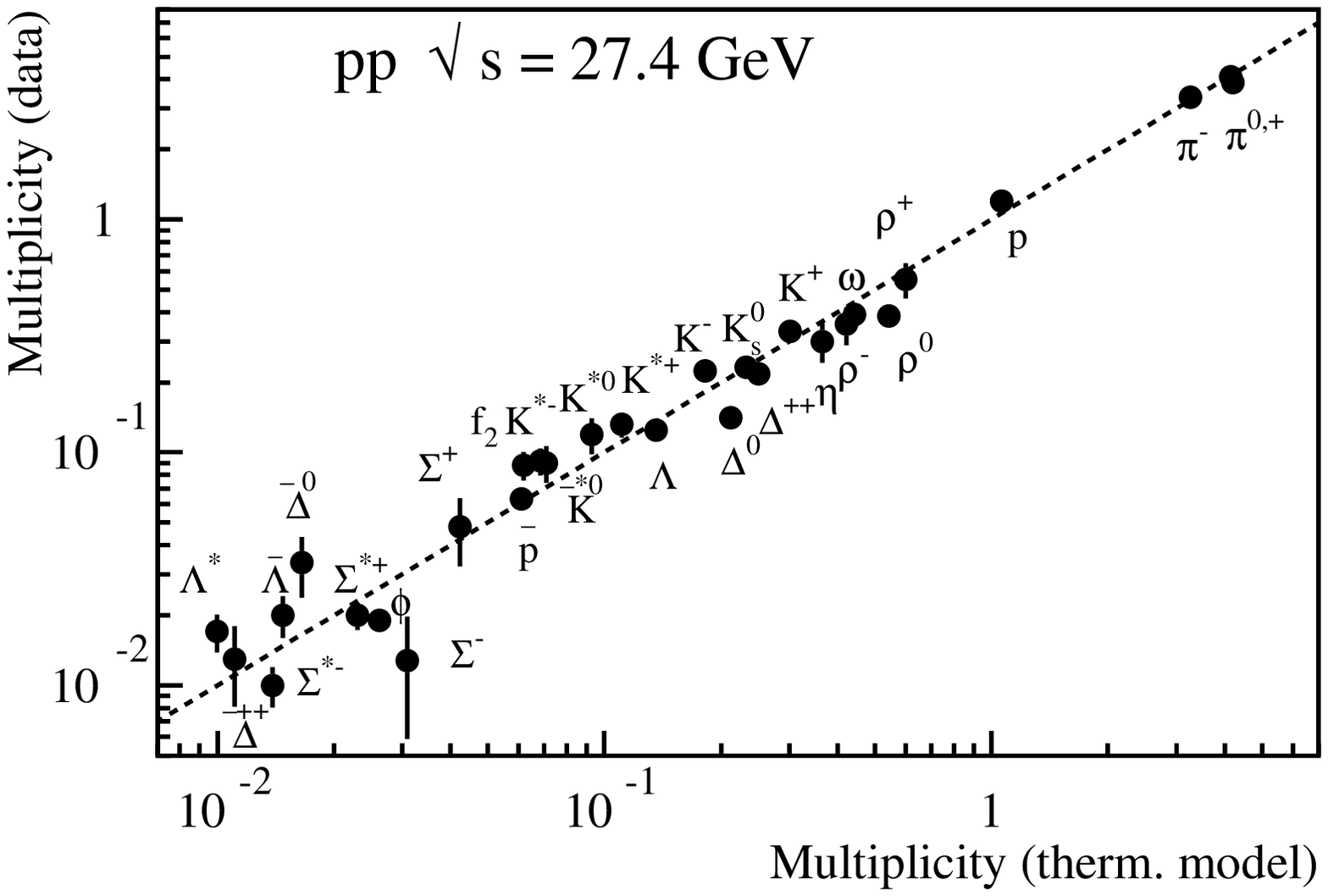,width=7.7cm, height=5cm}
\caption{Thermal hadronization in pp collisions, $T=175$ MeV 
\cite{Becattini}.}
\label{T-pp} 
\end{minipage}
\hspace{1.2cm}
\begin{minipage}[t]{7cm}
\vskip-4.8cm
%\hskip0.3cm 
\epsfig{file=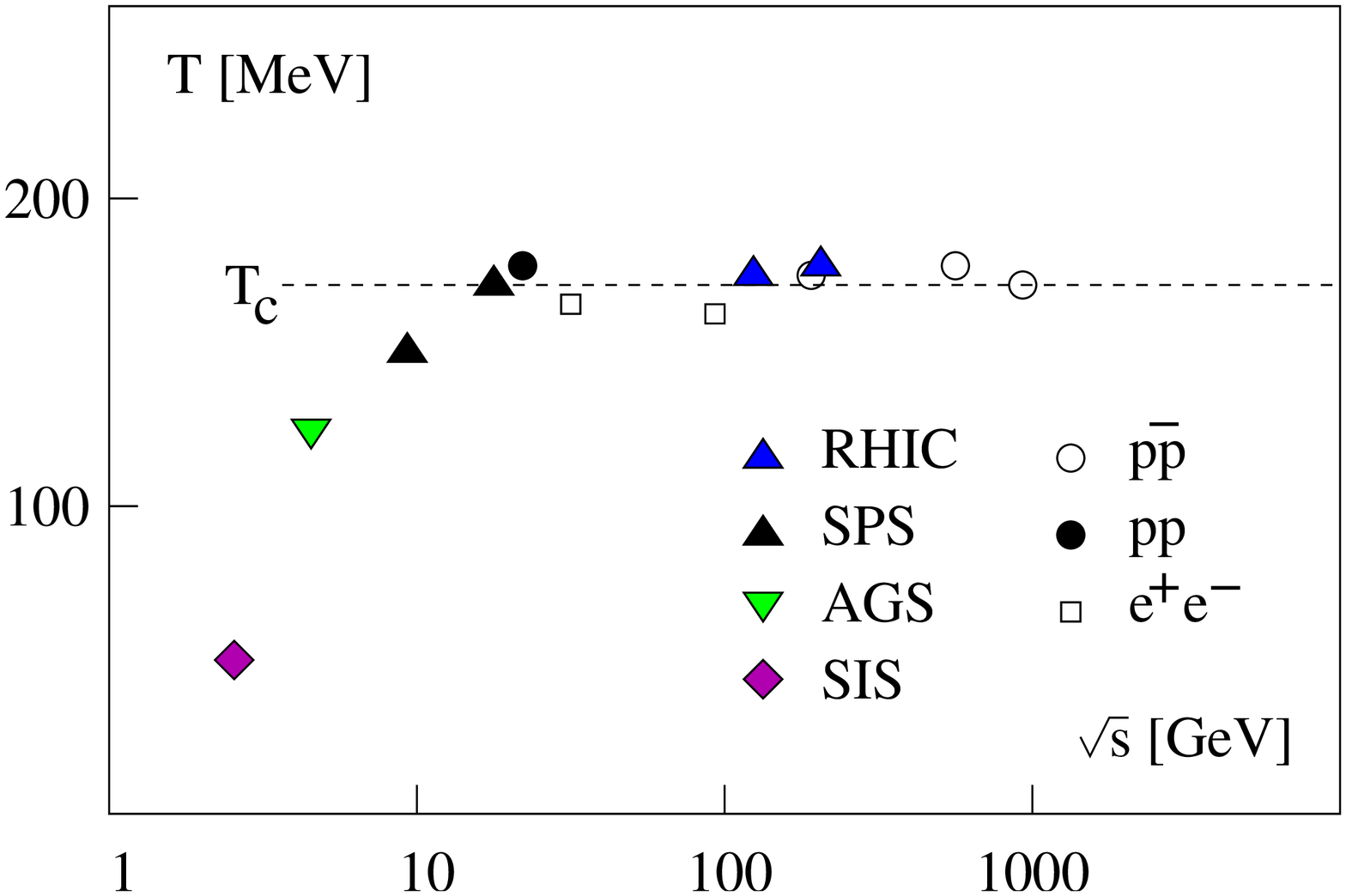,width=6.8cm, height=4.5cm}
\vskip0.3cm
\caption{Thermal hadronization temperatures.}
\label{T-E} 
\end{minipage}
\end{figure}

The partonic cascades produced by elementary hadron-hadron collision (or 
by $e^+e^-$ annihilation) evolve in space and time, eventually reaching 
a scale that requires hadronization. This transformation is known to 
lead to hadron abundances (for up to 30 different species) as given by 
an ideal resonance gas of a universal hadronization temperature $T_h$
\cite{Becattini}; see Fig.\ \ref{T-pp}. The hadronization temperature 
$T_h$ has been determined for elementary collisions at a variety of 
different collision energies (Fig.\ \ref{T-E}). It agrees well with 
that predicted for the QCD confinement transition (the line $T_c$ in 
Fig.\ \ref{T-E}), so that apparently the transition from partonic to 
hadronic degrees 
of freedom in such a system is in accord with statistical QCD. Since 
in the elementary collisions at the indicated $\sqrt s$, neither energy 
nor parton 
densities are high enough for any kind of deconfinement, the observed 
thermal abundance pattern is apparently not due to the formation of a 
deconfined medium in the pre-hadronization stage.   
 
\begin{figure}[htb]
%\vskip-1cm
\begin{minipage}[t]{7.5cm}
%\hskip-0.5cm
\epsfig{file=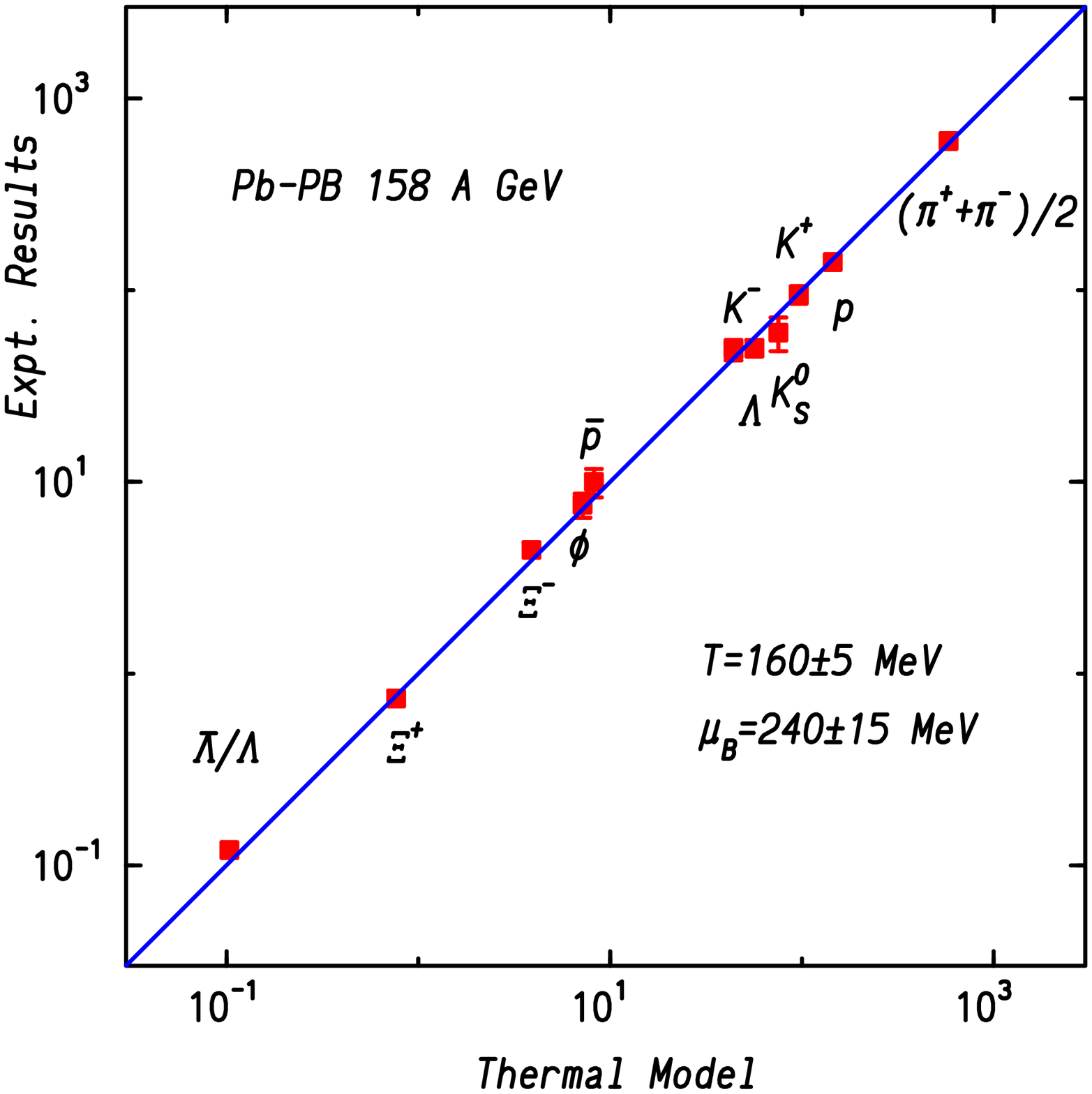,width=7cm, height=7.5cm}
\vskip-1cm
\caption{Hadron abundances in Pb-Pb collisions \cite{beca}.}
\label{T-Pb} 
\end{minipage}
\hspace{1cm}
\begin{minipage}[t]{7cm}
\vskip-7.5cm
%\hskip0.5cm
\epsfig{file=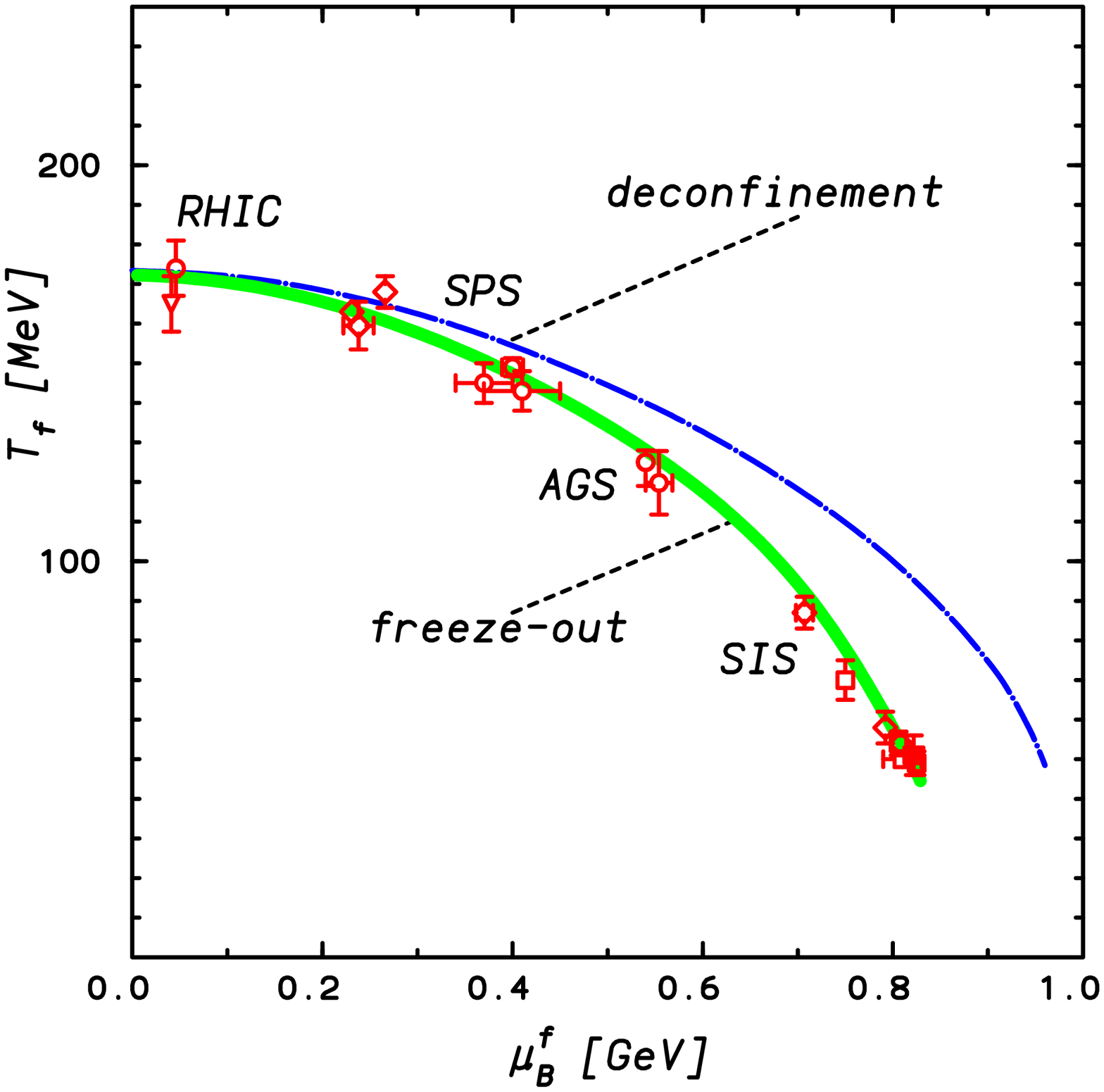,width=7cm, height=7.5cm}
\vskip-1cm
\caption{Freeze-out \cite{KR-FO}.}
\label{C-R} 
\end{minipage}
\end{figure}

\begin{figure}[htb]
%\hskip0.6cm
\begin{minipage}[t]{5cm}
\epsfig{file=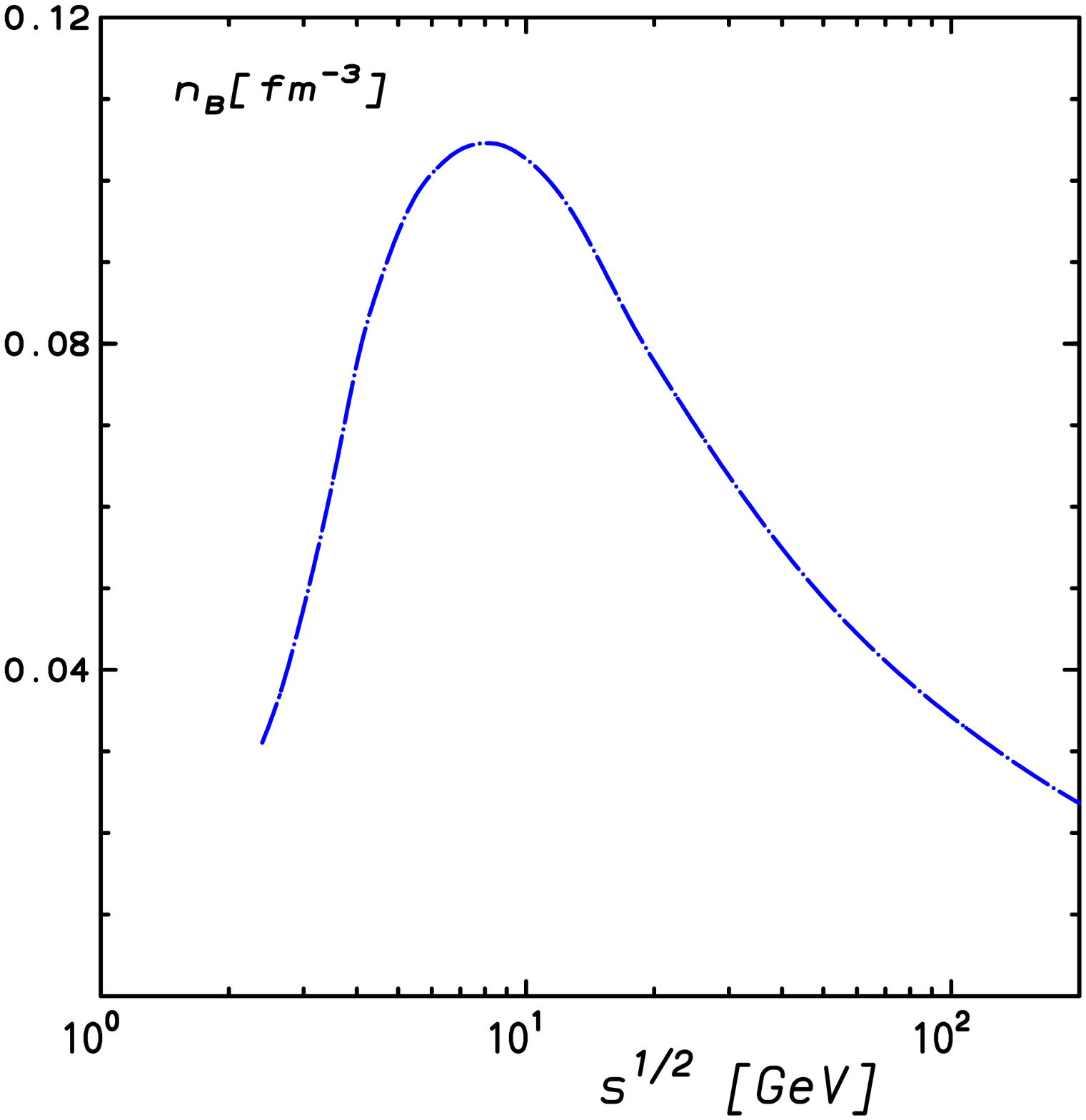,width=4.5cm, height=5.9cm}
\end{minipage}
\begin{minipage}[t]{5cm}
\vskip-5.26cm
\hskip-0.49cm
\epsfig{file=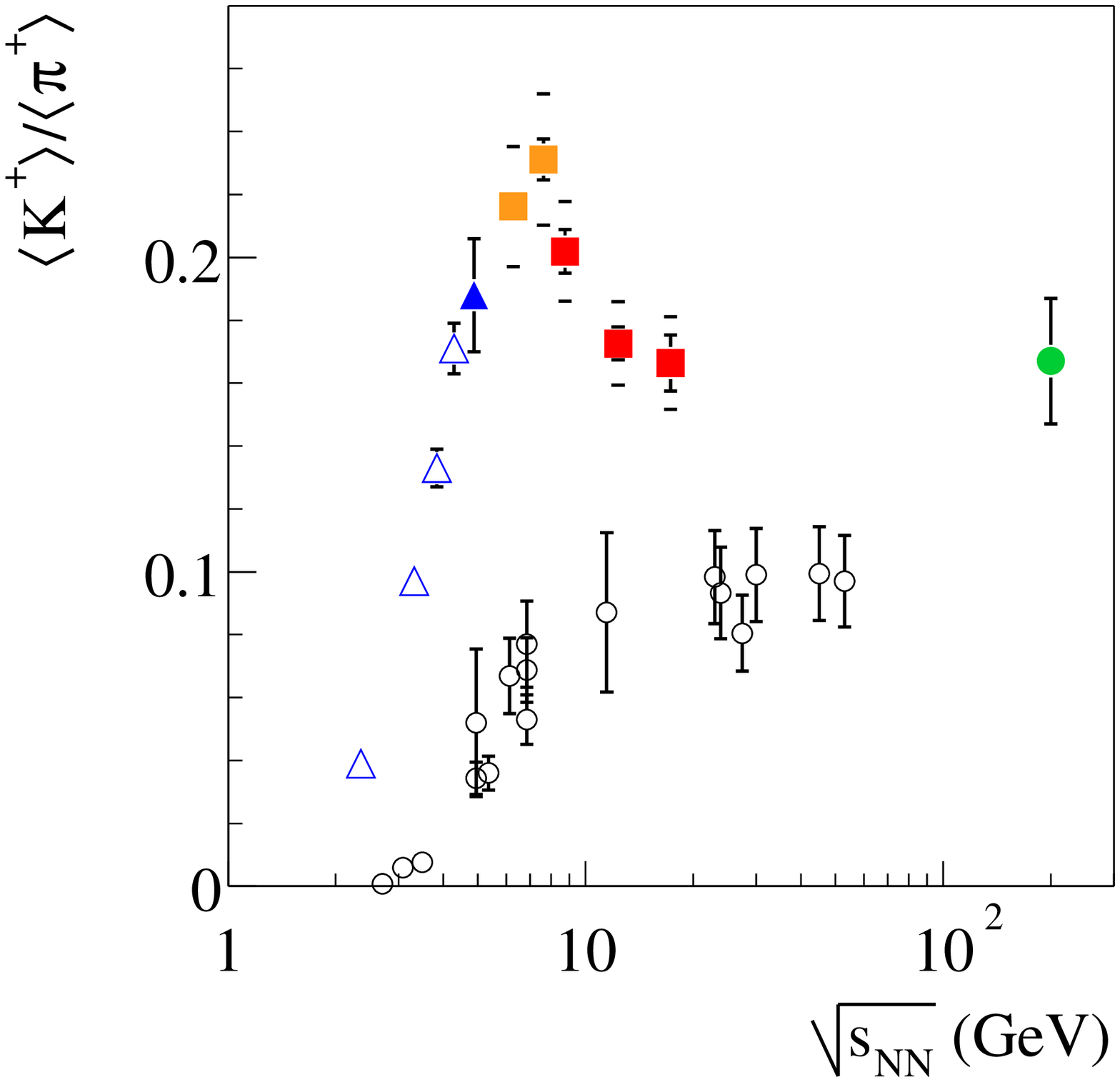,width=5.5cm, height=5.1cm}
\vskip-0.6cm
\end{minipage}
\hspace{0.15cm}
\begin{minipage}[t]{5cm}
\vskip-5.1cm
\epsfig{file=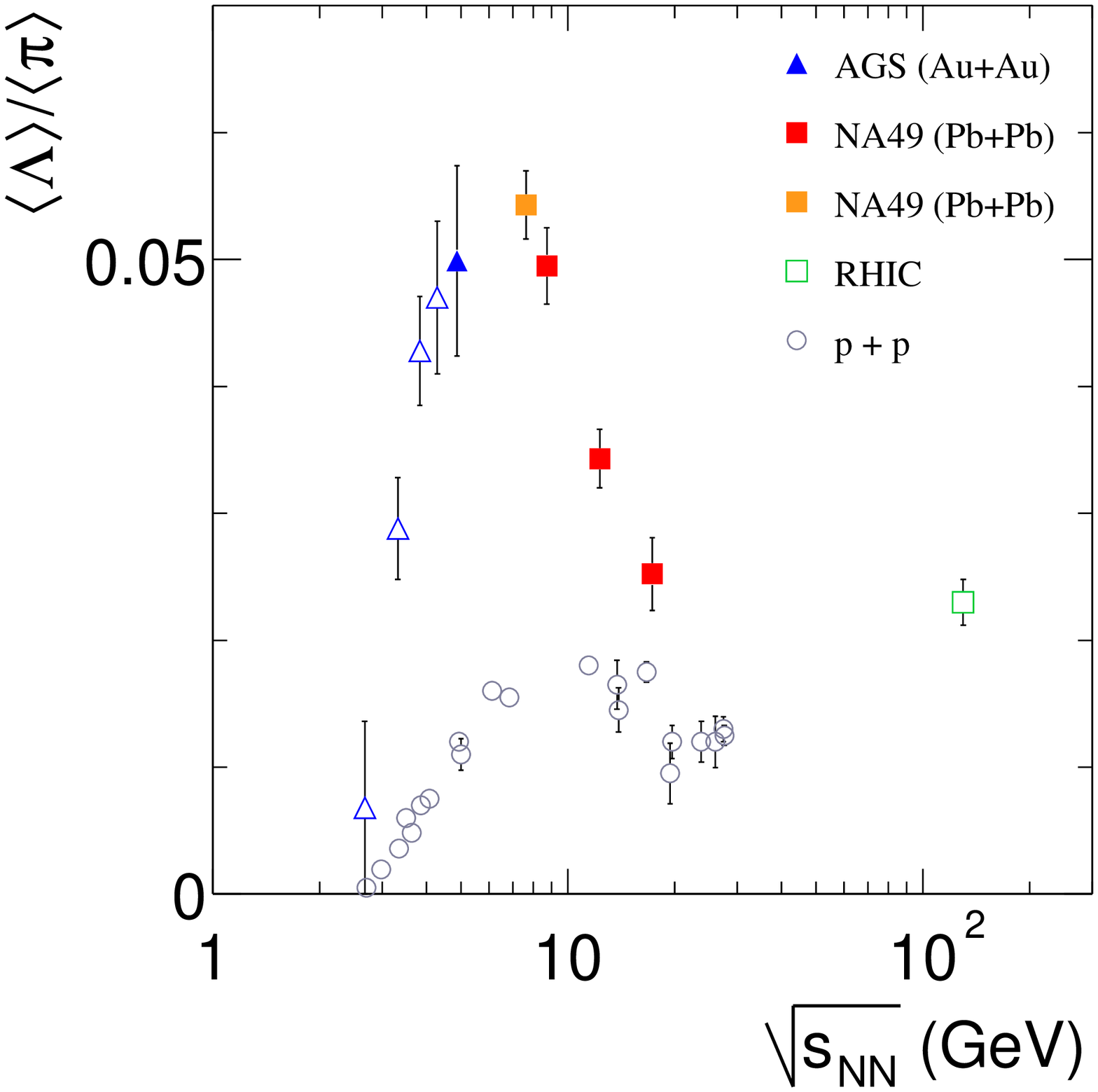,width=4.8cm, height=5cm,angle=-90}
\vskip-0.6cm
\end{minipage}
%\vspace{-0.3cm}

\vspace{-0.2cm}
~~\hskip2.7cm (a) \hskip4.5cm (b) \hskip 4.4cm (c)
\vspace{0.1cm}
\caption{Energy dependence at freeze-out for (a) baryon density \cite{KR-FO}, 
(b) $K^+/\pi^+$ and (c) $\Lambda/\pi^+$ \cite{Stock}.}
\label{KR}
\end{figure}

In central $AA$ collisions, nuclear stopping leads at AGS and SPS 
energies to a medium of a much higher baryon density than that formed 
in corresponding $pp$ interactions at mid-rapidity. Nuclear stopping is 
very much $\sqrt s$-dependent, which permits changes in the effective 
baryon density of the system under study. The basic question is whether 
the species abundances in $AA$ collisions are still those of an ideal 
resonance gas, with thermal parameters which register baryon density 
changes of the medium, both as function of $\sqrt s$ and in comparison 
to $pp$. The answer is clearly affirmative \cite{PBM}; we concentrate on 
central $AA$ collisions and always consider mid-rapidity, to avoid averaging
over different baryon densities.
\begin{itemize}
\vspace*{-0.2cm}
\item{From SIS to RHIC, the species abundances in $AA$ collisions are 
given by an ideal resonance gas, specified by a hadronization temperature 
$T$ and a baryochemical potential $\mu$; in Fig.\ \ref{T-Pb} we show the 
behaviour at the top SPS energy \cite{beca}. The resulting `freeze-out curve' 
in the $T,~\mu$  plane is shown in Fig.\ \ref{C-R} \cite{KR-FO}.}
\vspace*{-0.2cm}
\item{The baryon density in $AA$ collisions vanishes in the limit of 
high collision energy (`nuclear transparency'), so that $\mu \to 0$ for
$\sqrt s \to \infty$. The corresponding $T_h$ should therefore approach the 
deconfinement transition temperature $T_c$, as it does in high energy
elementary collisions. In Fig.\ \ref{T-E} it is seen that this is the case.}
\vspace*{-0.2cm}
\item{For $\mu \not=0$, the interacting hadron gas formed at the
confinement transition contains non-resonant baryon repulsion and
therefore cannot be approximated as an ideal resonance gas
\cite{Magas}. Hence for $\mu \not=0$, the freeze-out curve
no longer coincides with the deconfinement curve, as indicated in 
Fig.\ \ref{C-R}.}
\vspace*{-0.2cm}
\item{The baryon density $n_B(\sqrt s)$ along the freeze-out curve 
is shown in Fig.\ \ref{KR}a. With increasing $\sqrt s$, the two 
nuclei penetrate each other more and more, so that the baryon density 
increases. Around $\sqrt s \simeq 7$ GeV, the nuclei begin to pass 
through each other, nuclear transparency sets in, and $n_B$ starts to 
decrease.}
%\vspace*{-0.2cm}
\end{itemize}

\vskip-1cm

Thus nuclear stopping leads to a very characteristic $\sqrt s$-dependence 
of the baryon density along the freeze-out curve. Since the associated
production of strange particles increases with increasing baryon density, 
this behaviour is reflected in the ratios $K^+/\pi^+$ and $\Lambda/\pi$, 
as illustrated in Figs.\ \ref{KR}b and c \cite{Stock}. They are `enhanced' 
at finite $n_B$, and when $n_B \to 0$, they decreases to the finite value 
given by the resonance gas at $T=T_h,~\mu=0$. 

\medskip

It is thus evident that the hadronic medium in $AA$ collisions at 
freeze-out is a collective system whose properties can be 
accounted for by equilibrium thermodynamics, specified in terms of
a temperature $T$ and a baryochemical potential $\mu$. It is also clear
that a considerable part of the strangeness enhancement seen when
comparing $AA$ relative to $pp$ collisions, as in Fig.\ \ref{Antinori},
is due to the increase of baryon density of the medium seen by the 
hadronic probe. Thus it is necessary to determine what remains once 
this `normal strangeness enhancement' is taken into account. 

\medskip

The high energy behaviour of the ratios shown in Figs.\ \ref{KR}b and c
is just the $\mu \to 0$ limit. For both cases, the
corresponding results from $pp/p\bar p$ collsions are included.
We see that nuclear collisions provide at all energies species 
abundances in accord with hadronization through equilibrium thermodynamics;
in contrast, elementary hadron-hadron interactions lead to the mentioned 
`anomalous strangeness suppression'. Thus there is no strangeness 
enhancement in nuclear interactions; instead, we have to understand 
the observed deviations from thermal behaviour in elementary 
hadron-hadron collisions. 

\medskip 

The origin of this anomalous strangeness suppression is most likely
due to the small density of strange particles in elementary collisions.
This requires local strangeness conservation \cite{Hage-helium,Redlich}, 
which suppresses strange particle production. Only for the higher strange 
particle densities in $AA$ collisions, an ideal gas grand-canonical 
formulation becomes valid, removing the $pp$ suppression \cite{Redlich}.
At the same time, this shows that the medium provided by $AA$ collisions 
in the hadronic stage indeed shows large scale collective behaviour.

\medskip    

The low mass dilepton enhancement can be understood in terms of 
a modification of the $\rho$ in an interacting hadronic medium, 
changing its mass \cite{B-R}, its width \cite{Rapp}, or both. 
If this modification is primarily due to interactions of the $\rho$ 
with nucleons and nucleon resonances in a dense nuclear environment 
\cite{Rapp}, an increase of the baryon density should increase the effect. 
This is indeed observed \cite{Specht}, suggesting that also in this case 
the large baryon density of the system plays a crucial role for the 
observed difference between central $AA$ and $pp$ interactions. The
behaviour of the enhancement with increasing $\sqrt s$ will
clarify how much further thermal effects remain beyond this. 

\medskip

The energy dependence of hadronic probes in $AA$ collisions in the 
SPS range thus provides an excellent tool to investigate the baryon 
density dependence of hadronization and the resulting interacting 
hadron system. While many hadronic variables show a change of
behaviour when the baryon density starts to decrease, such studies 
have so far not revealed any clear threshold.

\medskip

The observed photon excess constitutes a first candidate for thermal
emission. Similarly, the intermediate mass dilepton enhancement has
been attributed to thermal emission during the evolution of the
system \cite{R-S}. In both cases, more data appear necessary to
identify the observed effects.

\medskip

It seems not easy to reconcile the observed HBT results with
an evolution of a hot thermal source. Increasing the collision 
energy will increase the initial energy density, and if the system
is thermal in an early deconfined stage, expansion should lead to
larger source sizes at higher energies. If the final source size
is defined by a freeze-out condition requiring a mean free path
of hadronic size \cite{Stock-A}, the baryonic composition of the 
hadronic medium can affect the source size, leading to a minimum 
of the freeze-out volume at the point of maximum baryon density 
\cite{PBM-source}. The expected increase would then occur only 
beyond this point, as the system becomes meson-dominated. A systematic 
high energy study in the RHIC/LHC range will certainly clarify this. 
Another, rather basic question is whether an individual $AA$ collision 
already produces a thermal medium, or whether only a superposition of 
many events leads to the observed thermal pattern. 

\medskip

The presence of elliptic flow can only be accounted for as 
consequence of the different pressure gradients in non-central 
interactions, and thus supports the presence of a thermal
medium on an event-by-event basis. It is striking, however, 
that the change of elliptic flow from out-of-plane low energy 
to in-plane high energy behaviour (see Fig.\ \ref{ellip}) occurs 
essentially at the turning point of the baryon density.
It thus remains to be clarified to what extent this effect
is due to the nuclear medium, and how much of it persists for 
$\mu \to 0$.

\medskip

The mass-dependent broadening of hadronic transverse momentum 
spectra is in accord with predictions from radial flow studies,
which also assume a thermal medium per single event. Here, however, 
the role of initial state effects due to production from 
multiple scattering is not fully clarified. Hydrodynamic radial flow of
an expanding thermal source would in general lead to a further
increase of broadening with collision energy, while initial state
effects would result in saturation. Hence a comparison between
SPS and higher energy data should resolve the issue.

\medskip

In summary:

\begin{itemize}
\vspace*{-0.2cm}
\item{The centrality dependence of \J~production shows a clear onset of 
`anomalous' behaviour, indicating the formation of a deconfined partonic 
medium.}
\vspace*{-0.2cm}
\item{Species abundances and strangeness production in high
energy $AA$ collisions are in accord with 
ideal resonance gas thermodynamics 
at the confinement/deconfinement transition.}
\vspace*{-0.1cm}
\end{itemize}

\noindent
Our views today thus are clearer, but also somewhat different
from what they were at the beginning of the programme. We have evidence
for deconfinement as well as for thermal behaviour, but at different 
evolution stages. What we know now
is largely due to the pioneering work of the SPS experiments. It is 
evident that we need more work to really reach final conclusions. It is 
also  evident, however, that the search for critical behaviour in nuclear 
collisions, on the partonic as well as on the hadronic side, 
requires looking for onsets, and here the SPS, 
with its energy range and the statistics bonus of a fixed target machine, 
is unique. Much of the further work needed will be done at the SPS, or 
it will not be done at all.

\vskip1cm

\centerline{\large{\bf {Acknowledgements}}}
    
\bigskip

It is a pleasure to thank many colleagues of the SPS Heavy Ion 
Programme for innumerable stimulating discussions. Particular thanks 
for help in the preparation of this report go to P.\ Braun-Munzinger, 
F.\ Karsch, C.\ Louren{\c c}o, K.\ Redlich, H.-J.\ Specht, J.\ Stachel 
and R.\ Stock.


\begin{thebibliography}{99}

\bibitem{karsch} For a recent review, see 
F.\ Karsch and E.\ Laermann, {\sl Quark-Gluon Plasma 3},\\
R.\ C.\ Hwa and X.-N.\ Wang (Eds.), World Scientific 2004, 1.

\bibitem{Celik} T.\ {\c C}elik, J.\ Engels and H.\ Satz, \PL 129 B (1983) 323.

\bibitem{Biele1} J.\ Engels et al., \PL 101B (1981) 89. 

\bibitem{kuti} L.\ D.\ McLerran and B.\ Svetitsky, \PL 98B (1981) 195;\\
J.\ Kuti et al., \PL 98B (1981) 199.

\bibitem{Biele2} F.\ Karsch and E.\ Laermann, \PR D 50 (1994) 6954.

\bibitem{Fodor} See e.g.\ Z.\ Fodor, \NP A715 (2003) 319c.

\bibitem{bar-fluc} C.\ R.\ Allton et al., \PR D 68 (2003) 014507.

\bibitem{Bjorken83} J.D. Bjorken, Phys. Rev. D27 (1983) 140.

\bibitem{Shuryak} E.V. Shuryak, Phys. Rep. 61 (1980) 71.

\bibitem{Keijo} K. Kajantie and H.I. Miettinen, Z. Phys. C 9 (1981) 341.

\bibitem{MS} T. Matsui and H. Satz, Phys. Lett. B178 (1986) 416.

\bibitem{jets} J.D. Bjorken, Fermilab-Pub-82/59-THY (1982) and Erratum.

\bibitem{rho-chiral} R. Pisarski, Phys. Lett. B110 (1982) 155.

\bibitem{flow} L. Van Hove, Phys. Lett. B118 (1982) 138.

\bibitem{HBT} M. Gyulassy et al., Phys. Rev. C20 (1979) 2267.

\bibitem{Stock-A} R. Stock, Annalen Phys. 48 (1991) 195. 

\bibitem{Hagedorn} R.\ Hagedorn, Nuovo Cim. Suppl. 3 (1965) 147;\\
Nuovo Cim. 56A (1968) 1027.

\bibitem{Becattini} F.\ Becattini, \ZP C 69 (1996) 485;\\
F.\ Becattini and U.\ Heinz, \ZP C 76 (1997) 269;\\
F.\ Becattini et al., \PR C 64 (2001) 024901.

\bibitem{Rafelski} J.\ Rafelski and B.\ M\"uller, 
Phys. Rev. Lett. 48 (1982) 1066.

\bibitem{strangelet} S.\ A.\ Chin and A.\ K.\ Kerman, \PRL 43 (1979);\\
E.\ Witten, \PR D30 (1984) 272.

\bibitem{NA50} M.C. Abreu et al. (NA50), Phys. Lett. B410 (1997) 337;\\
M.C. Abreu et al. (NA50), Phys. Lett. B450 (1999) 456;\\
M.C. Abreu et al. (NA50), Phys. Lett. B477 (2000) 28.

\bibitem{NA38/50} C. Baglin et al. (NA38), Phys. Lett. B220 (1989) 471;\\
M.C. Abreu et al. (NA50), Phys. Lett. B410 (1997) 337.

\bibitem{Ceres} G. Agakichiev et al. (CERES), Phys. Rev. Lett. 75 (1995) 
1272;\\
G. Agakichiev et al. (CERES), Phys. Lett. B422 (1998) 405;\\
B. Lenkeit et al. (CERES), Nucl. Phys.  A661 (1999) 23c;\\
J. P. Wessels et al. (CERES), \NP A715 (2003) 262c.

\bibitem{Masera} N.\ Masera et al. (HELIOS-3), Nucl. Phys. A590 (1995) 93c.

\bibitem{Specht} D.\ Adamov\'a et al. (CERES), \PRL 91 (2003) 042301.

\bibitem{photondata} M.M. Aggarwal et al. (WA98), 
Phys. Rev. Lett. 85 (2000) 3595.

\bibitem{hm-dileptons} E. Scomparin et al. (NA50), Nucl. Phys. A610 
(1996) 331c.

\bibitem{stock} R.\ Stock, Report at QM2004, Oakland/California, USA

%\bibitem{Appel} H. Appelsh\"auser et al. (CERES), Nucl. Phys. A698 (2002) 253.

%\bibitem{adcox} Compiled in K. Adcox et al. (PHENIX), Phys. Rev. Lett. 88 
%(2002) 192302;\\
%see also D.\ Adamov\'a et al.\ (CERES), \NP A 714 (2003) 124.

\bibitem{antinori} K. Fanebust et al. (NA57), J. Phys. G 28 (2002) 1607;\\
V.\ Manzari et al.\ (NA57), \NP A715 (2003) 140c.

\bibitem{newmass} M.\ Weber et al. (NA52), J.\ Phys.\ G28 (2002) 1921.

\bibitem{PP} N.\ Armesto et al., \PRL 77 (1996) 3736;\\
M.\ Nardi and H.\ Satz, \PL B 442 (1998) 14;\\
H.\ Satz, \NP A 661 (1999) 104c.

\bibitem{cgc} L.\ McLerran and R.\ Venugopalan, \PR D 49 (1994) 2233
and 3352; \\
D 50 (1994) 2225;
for a recent review, see \\
L.\ McLerran, \NP A 702 (2002) 49.

\bibitem{spectral} M.\ Asakawa and T.\ Hatsuda, \PRL 92 (2004) 012001;\\
P.\ Petreczky et al., \NP B 129\&130 (2004) 596.

\bibitem{DFS} S.\ Digal, S.\ Fortunato and H.\ Satz, \EP~32 (2004) 547.

\bibitem{Ramona} R.\ Vogt, Phys.\ Rept.\ 310 (1999) 197.

\bibitem{b-o} J.-P. Blaizot, M.\ Dinh and J.-Y.\ Ollitrault, 
\PRL 85 (2000) 4012.

\bibitem{PBM} P.\ Braun-Munzinger and J.\ Stachel, \NP A 606 (1996) 320 and\\
\NP A 638 (1998) 3c.

\bibitem{beca} Analysis of NA49 data by
F.\ Becattini et al.,\PR C 64 (2001) 024901;\\
for a more recent analysis, see \\
P.\ Braun-Munzinger, K.\ Redlich and
J.\ Stachel, {\sl Quark-Gluon Plasma 3}, \\
R.\ C.\ Hwa and X.-N.\ Wang
(Eds.), World Scientific Publ., Singapore 2004.

\bibitem{KR-FO}J.\ Cleymans and K.\ Redlich, \PRL 81 (1998) 5284.

\bibitem{Magas} V.\ Magas and H.\ Satz, \EP ~32 (2003) 115.

\bibitem{Stock} Compiled by R.\ Stock, hep-ph/0404125. 

\bibitem{Hage-helium} R.\ Hagedorn, CERN Rept. 71 (1971);\\
E.\ V.\ Shuryak, \PL B42 (1972) 357.  

\bibitem{Redlich} S.\ Hamieh, K.\ Redlich and A.\ Tounsi, 
\PL B 486 (2000) 61. 

\bibitem{B-R} G.\ E.\ Brown and M.\ Rho, Phys.\ Rept.\ 363 (2002) 85. 

\bibitem{Rapp} R.\ Rapp and J.\ Wambach, Adv.\ Nucl.\ Phys.\ 25 (2000) 1.

\bibitem{R-S} R. Rapp and E. Shuryak, Phys. Lett. B473 (2000) 13.

\bibitem{PBM-source} D.\ Adamov\'a et al.\ (CERES), \PRL 90 (2003) 022301.


\end{thebibliography}
\end{document}